\documentclass[preprint,11pt,a4paper]{elsarticle}
\usepackage{lineno}
\usepackage{xcolor}
\definecolor{revisionred}{RGB}{0,0,0}

\newcommand{\updated}[1]{#1}
\newcommand{\updatedfigure}[1]{#1}
\newcommand{\newrevision}[1]{#1}
\newcommand{\newrevisionfigure}[1]{#1}

\usepackage{etoolbox}
\usepackage[a4paper,left=22mm,right=22mm,top=25mm,bottom=25mm]{geometry}
\usepackage{setspace}
\usepackage[T1]{fontenc}
\usepackage[utf8]{inputenc}
\usepackage{newtxtext,newtxmath}
\usepackage{microtype}

\usepackage{amsmath,amssymb,booktabs,graphicx,array}
\usepackage[font=small,labelfont=bf,labelsep=period]{caption}
\usepackage{tikz}
\usetikzlibrary{arrows.meta,positioning,fit,shapes.geometric}
\usepackage{url}
\usepackage[colorlinks=true,
            linkcolor=blue,   
            citecolor=blue,   
            urlcolor=blue]{hyperref}
\usepackage{lastpage}
\usepackage{placeins}

\biboptions{authoryear}

\makeatletter
\renewcommand\section{\@startsection{section}{1}{\z@}%
  {18pt plus 3pt minus 2pt}{8pt plus 1pt}{\normalfont\large\bfseries}}
\renewcommand\subsection{\@startsection{subsection}{2}{\z@}%
  {12pt plus 2pt minus 1pt}{6pt plus 1pt}{\normalfont\normalsize\bfseries}}
\renewcommand\subsubsection{\@startsection{subsubsection}{3}{\z@}%
  {10pt plus 2pt minus 1pt}{4pt plus 1pt}{\normalfont\normalsize\itshape}}
\apptocmd{\ps@pprintTitle}{%
  \let\paper@titlefoot\@oddfoot
  \def\@oddfoot{\paper@titlefoot\unskip}%
  \let\@evenfoot\@oddfoot
}{}{}
\makeatother

\newcommand{\papertitle}{Diary matching and the spatial distribution of synthetic travel demand: behavioural representation learning within eqasim}

\newcommand{\paperauthors}{%
\author[inst1]{Abdelkader Dib}
\ead{dabdelkader@ethz.ch}
\author[inst1]{Milos Balac}
\address[inst1]{Center for Sustainable Future Mobility (CSFM), ETH Zürich, Switzerland}%
}

\newcommand{\EnergyGain}{\updated{19.5}}
\newcommand{\ChainGain}{\updated{18.5}}
\newcommand{\DistanceGain}{\updated{19.4}}

\journal{Transportation}

\begin{document}
\raggedbottom

\begin{frontmatter}

\title{\papertitle}
\paperauthors

\begin{abstract}
The assignment of daily plans to synthetic individuals largely determines the travel demand used in transport planning and policy evaluation. Matching procedures based on categorical attributes may separate individuals with similar travel behaviour at the boundaries of age or income classes, and agreement at the aggregate level may conceal differences in local demand patterns. We propose Behaviour-Supervised Latent Matching (BSLM), an encoder--decoder approach that learns a 16-dimensional representation of personal, household and residential attributes, supervised by observed travel behaviour. Complete survey diaries are then transferred to synthetic individuals from nearby respondents in this latent space. We compare BSLM with the default Recursive Hot-Deck Matching (RHDM) of eqasim and with an order-independent categorical baseline. In a replay experiment on the Swiss Microcensus, using 37,876 respondents, three donor-exclusion groups and 100 diary draws per person, BSLM reduces the joint-behaviour, ordered-chain and daily-distance scores by \EnergyGain\%, \ChainGain\% and \DistanceGain\%, respectively, relative to RHDM. We further assign diaries to 850,663 synthetic residents in twenty independent runs. National trip counts are nearly identical across methods, but local purpose patterns differ: shopping trips increase by 0.6\% nationally with BSLM compared to RHDM, yet decrease by 5.5\% among residents of central Geneva, with larger differences across residential cells. Across twenty random seeds, city-centre mean daily distances by purpose show min--max spans reaching 28.7\% of their twenty-run mean. Diary matching should therefore be assessed with individual-level diagnostics and local sensitivity analyses, in addition to aggregate agreement, before its results are used to inform planning decisions.
\end{abstract}


\end{frontmatter}

\section{Introduction}
\label{sec:introduction}
Daily plans link a synthetic population to the travel demand used in agent-based and activity-based transport models. While a person record describes an individual and the associated household, a daily plan additionally specifies the activities performed during the day, as well as the trips and modes connecting them, including their timing. The assignment of daily plans to the synthetic population therefore determines who generates travel, for which purposes, and at what times of day. This is relevant for transport planning, since two demand scenarios with similar aggregate trip totals may imply substantially different requirements for public transport capacity, access to employment, or access to local services.

Beyond these planning uses, agent-based simulations are increasingly applied to a broad set of impact assessments built on the same synthetic demand. These include pollutant emissions \citep{dib2025microscopicEmissions,dib2027mesoscopicEmissions}, population exposure to these emissions \citep{kickhofer2015pricing,agarwal2020onroad}, noise exposure \citep{kaddoura2017noise,lebescond2023population}, accessibility \citep{ilahi2017accessibility,efthymiou2013multidimensional}, and the distributional effects of policies across population groups \citep{kickhofer2011income,meyerdefreitas2017equity}. Such indicators depend not only on the total amount of travel, but on who travels, where, and when: noise and emissions are sensitive to the timing and spatial distribution of trips, exposure depends on the locations and durations of individual activities, and distributional analyses depend on the sociodemographic attributes attached to each plan. If daily plans are assigned to synthetic persons in a way that does not preserve these relationships, the resulting errors are not confined to the demand model. They propagate to all downstream indicators and may change the conclusions drawn from the simulation, including the relative ranking of the policies under evaluation.

The eqasim framework provides a modular process for generating synthetic population and preparing transport simulations \citep{horl2021eqasim}. Its applications include Paris and \^{I}le-de-France \citep{horl2021paris}, S\~ao Paulo \citep{sallard2021saopaulo}, California \citep{balac2021california} and Switzerland \citep{switzerlandScenario}. The established default diary-assignment approach in eqasim is Recursive Hot-Deck Matching (RHDM): a synthetic person receives an observed survey day from a donor selected using categorical characteristics and an ordered relaxation procedure \citep{eqasimHotDeck}. This keeps the components of an observed day together and fits the subsequent location-assignment and simulation stages.

The main challenge is to identify survey respondents whose behaviour is suitable for each synthetic person. Fixed age or income categories are clear and easy to use, but their boundaries do not necessarily mark changes in travel behaviour. Two people on opposite sides of a boundary may have similar days, while people in the same category may differ in household responsibilities, car access or home setting. Adding attributes creates smaller donor groups, and some may contain few or no respondents. Relaxing the attributes solves this shortage, but the relaxation order affects which differences are accepted first.

We propose Behaviour-Supervised Latent Matching (BSLM), a method for attaching
observed travel diaries to synthetic agents. BSLM learns a 16-dimensional
representation of personal, household and residential characteristics by
training a model to predict several features of the observed travel diaries.
All attributes are used at their original level of detail, without prior
aggregation into coarse classes. Similarity in the learned representation
then defines the set of candidate survey donors for each synthetic agent.
The method transfers a complete observed day from the selected donor to the
synthetic agent, instead of generating a new sequence of activities. This
choice preserves the existing eqasim workflow and avoids introducing a
separate schedule generation and consistency checking step.

We present three experiments in sequence. First, a Microcensus replay tests how well each method assigns another respondent's day when the recipient's own household is excluded from the donor pool. Second, repeated matching of the Swiss synthetic population compares national and canton-level travel patterns. Third, a local analysis examines residents of central Z\"urich and Geneva and the variation across residential cells. Each experiment is described briefly before its results. We also compare an Order-independent Hot-Deck Matching (OHDM) method, which changes categorical relaxation without learning a representation.

The contribution is both methodological and relevant to planning practice. We describe how behavioural supervision can guide whole-diary matching, evaluate its replay performance using distributional scores, and show how differences between matching methods vary with spatial aggregation. The population analysis is not a test against known true demand. Instead, it examines whether apparent national agreement is sufficient to regard alternative diary-assignment methods as interchangeable for local planning.

\section{Background and related work}
\label{sec:background}

Population synthesis and diary assignment address related but distinct problems. Synthesis generates a population of persons and households whose demographic characteristics are consistent with census data, whereas diary assignment equips these persons with daily activity plans, i.e., the activities they perform and the trips connecting them. The eqasim pipeline combines both steps in a modular, open-data workflow in which census and household travel survey data are processed into a synthetic travel demand, as demonstrated for Paris and Île-de-France \citep{horl2021paris}. Other frameworks, such as mobiTopp, likewise connect synthetic persons with activity schedules and behavioural simulation \citep{mallig2013mobitopp}.

In practice, eqasim is most commonly used to provide travel demand for MATSim, an agent-based transport simulation framework \citep{horni2016matsim}. In MATSim, each agent is described by a daily plan. All plans are executed simultaneously in a mobility simulation on the transport network and are iteratively improved through replanning until the system approaches a stochastic user equilibrium. More recently, eqasim-based demand has also been used outside of MATSim, for instance as input to the microscopic simulator SUMO \citep{ngari2026sumo}. Regardless of the downstream simulator, the initial diary assignment defines the demand supplied to all later modelling stages. A change in this assignment can therefore alter simulation outcomes even when the underlying synthetic population remains fixed.

Hot-deck imputation copies observed information from suitable donors. Donor availability, the variables used to define similarity and the selection rule all affect the result \citep{andridge2010hotdeck}. For travel diaries, copying an entire day preserves observed combinations of purposes, modes and timing. However, a day that is coherent for its original respondent is not necessarily appropriate for every recipient. Categorical matching makes this suitability explicit through matching cells, but introduces boundaries and a priority order when those cells are relaxed.

Representation learning offers a different way to define similarity. An encoder maps many inputs to a smaller vector; an autoencoder also reconstructs inputs from that vector \citep{hinton2006reduction,goodfellow2016deep}. Multi-task learning shapes a shared representation through several prediction targets \citep{caruana1997multitask}. BSLM combines these ideas: predicting travel outcomes gives the representation a behavioural purpose, while input reconstruction helps retain personal and household information. The learned vector is used for donor retrieval, not direct schedule generation.

Generative models provide a related alternative. Deep generative approaches have been applied to synthetic people \citep{borysov2019microagents} and vehicle stop trajectories \citep{chiesa2022traffic}. They can produce combinations absent from the training data. Matching instead remains within the observed diary library. More directly for activity schedules, \citet{shone2025schedules} use variational autoencoders to synthesise activity participation and duration, together with an evaluation framework for generated schedules. Our model uses behavioural prediction to select observed days, so it addresses a different output task and is not a direct benchmark against their generator. We adopt the latter approach to preserve the existing eqasim structure and retain an inspectable observed day for each assignment. The corresponding limitation is that no matching method can recover behaviour that is entirely absent from its donors. The approach therefore requires a representative donor dataset.

Survey diaries can also complement emerging mobility data. \citet{tozluoglu2026plans} combine mobile-phone application records with survey information to construct variable weekday plans in Sweden. Their evaluation considers activity sequences, temporal patterns and trip distances. This provides a related transport-planning context for combining person-level schedule construction with several behavioural diagnostics, although its data and modelling task differ from synthetic-person matching.

Validation must also distinguish suitable assignments from plausible totals. Aggregate activity-chain frequencies are an established diagnostic in travel-demand synthesis \citep{horl2021paris}, but do not identify who receives each chain. Proper scoring rules assess a distribution of possible days against an observed day \citep{gneiting2007scores}. They can therefore complement chain-frequency plots, activity counts and distance distributions. Our analysis uses both approaches, while distinguishing survey replay from the application to a population whose individual days are unobserved.

\section{Matching methods}
\label{sec:methods}
\subsection{The task and categorical comparators}
Let $\mathcal D=\{(x_j,y_j,w_j)\}_{j=1}^n$ denote survey respondents with input characteristics $x_j$, observed diaries $y_j$ and positive survey weights $w_j$. A synthetic recipient has characteristics $x_i^\star$ but no observed diary. A matcher selects a donor $J_i$ and assigns their complete day $y_{J_i}$. Repeating that selection produces a distribution of possible days. The aim is not to predict one uniquely correct schedule, but to attach reasonable behavioural variation to people with different characteristics.

\begin{figure}[h]
    \centering
    \includegraphics[width=0.6\linewidth]{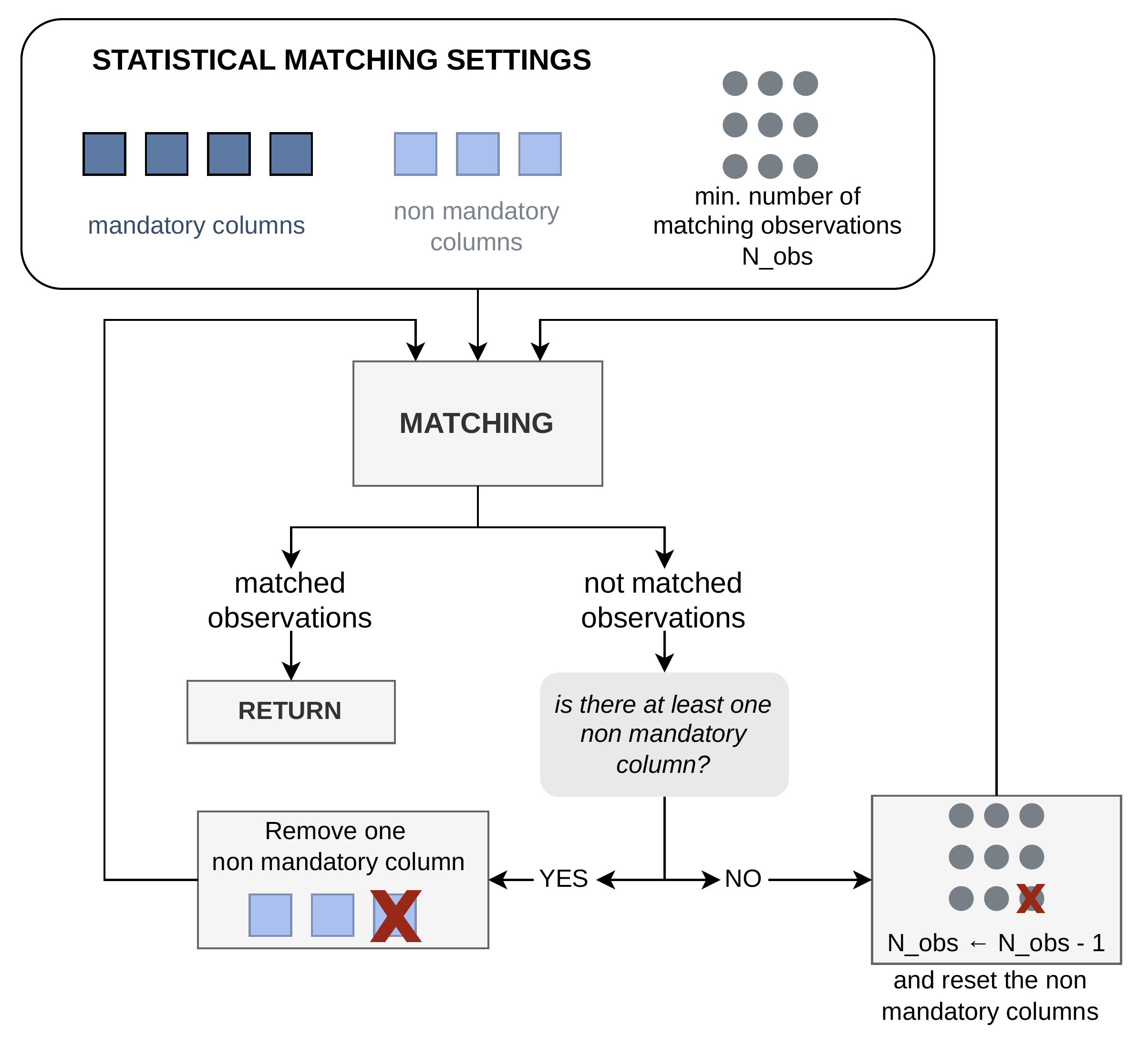}
    \caption{Graphical representation of the statistical matching algorithm implementation \citep{switzerlandScenario}.}
    \label{fig:rhdm_matching}
\end{figure}

RHDM, the default method in eqasim, divides matching attributes into mandatory and non-mandatory groups (\autoref{fig:rhdm_matching}). Mandatory attributes have priority: they are required to agree during the main search. Non-mandatory attributes are preferences that can be removed in a fixed order when there are too few donors. The search starts with all attributes and a preferred pool of ten donors. If this is unsuccessful, it removes the lowest-priority non-mandatory attribute and tries again. Once those attributes are exhausted, it reduces the required donor count and restarts the search. In the final fallback, even mandatory attributes can be relaxed. Thus, ``mandatory'' describes their priority in the main search, not an absolute constraint in every fallback.

For adults in this study, the mandatory group includes age group, sex, car availability, employment, public-transport quality, density, children, residential region and broad commuting-distance group. The non-mandatory group includes finer commuting-distance groups, income, canton, workplace arrangement and urban setting. Age-specific profiles are used for children and young adults. Within the accepted donor pool, selection probabilities are proportional to survey weights \citep{andridge2010hotdeck}. \autoref{fig:rhdm_matching} illustrates the search; the study settings follow the Switzerland application \citep{switzerlandScenario}.

OHDM also distinguishes mandatory and non-mandatory matching attributes, but compares the number of disagreements instead of removing attributes in a fixed order. It first applies hard compatibility constraints, which are never relaxed. Within this feasible pool, it keeps exact mandatory matches whenever any exist. Only when none exist does it keep donors with the smallest possible number of mandatory disagreements. It then adds non-mandatory disagreements in levels: first none, then at most one, then at most two, and so on, until the pool reaches ten donors or no further candidates can be added. All combinations at a given disagreement level are treated alike. This removes the priority between individual non-mandatory attributes.

\begin{figure}[tb]
\centering
\resizebox{.7\textwidth}{!}{%
\begin{tikzpicture}[>=Latex,
 box/.style={draw,rounded corners,fill=gray!4,align=center,font=\small,text width=110mm,minimum height=9mm},
 decision/.style={diamond,aspect=2.5,draw,fill=blue!4,align=center,font=\small,text width=26mm,inner sep=1mm},
 action/.style={draw,rounded corners,align=center,font=\small,text width=40mm,minimum height=12mm},
 flow/.style={thick,->},line/.style={thick}]
\node[box] (settings) at (0,0) {Hard constraints, mandatory preferences and non-mandatory preferences\\Preferred pool: 10 donors};
\node[box] (hard) at (0,-1.35) {Keep donors satisfying the hard constraints};
\node[box] (mandatory) at (0,-2.7) {Keep exact mandatory matches if available; otherwise keep\\the donors with the fewest mandatory disagreements};
\node[box] (pool) at (0,-4.1) {Keep donors with at most $r$ non-mandatory disagreements\\Start with $r=0$};
\node[decision] (enough) at (-2.6,-6.3) {At least\\10 donors?};
\node[decision] (more) at (3.6,-6.3) {More disagreement\\levels available?};
\coordinate (merge) at (-2.6,-7.7);
\node[action,fill=green!5,text width=50mm] (sample) at (-2.6,-8.8) {Survey-weighted donor draw};
\node[action,fill=blue!4] (expand) at (3.6,-8.8) {Increase $r$ by one};
\draw[flow] (settings) -- (hard);
\draw[flow] (hard) -- (mandatory);
\draw[flow] (mandatory) -- (pool);
\draw[flow] (pool.south) -- (0,-4.9) -| (enough.north);
\draw[flow] (enough.east) -- node[above,font=\small] {No} (more.west);
\draw[line] (enough.south) -- node[left,font=\small] {Yes} (merge);
\draw[line] (more.south west) |- node[pos=.7,above,font=\small] {No} (merge);
\fill (merge) circle (1.4pt);
\draw[flow] (merge) -- (sample.north);
\draw[flow] (more.south) -- node[right,font=\small] {Yes} (expand.north);
\draw[flow] (expand.east) -- (6.9,-8.8) -- (6.9,-4.05) -- (pool.east);
\end{tikzpicture}}
\caption{OHDM donor search. The candidate pool expands by the number of non-mandatory disagreements, without a fixed order between individual attributes. Mandatory preferences are relaxed only when exact agreement is unavailable. Hard constraints remain fixed; if no donor satisfies them, the recipient is reported as unmatched.}
\label{fig:ohdm_matching}
\end{figure}
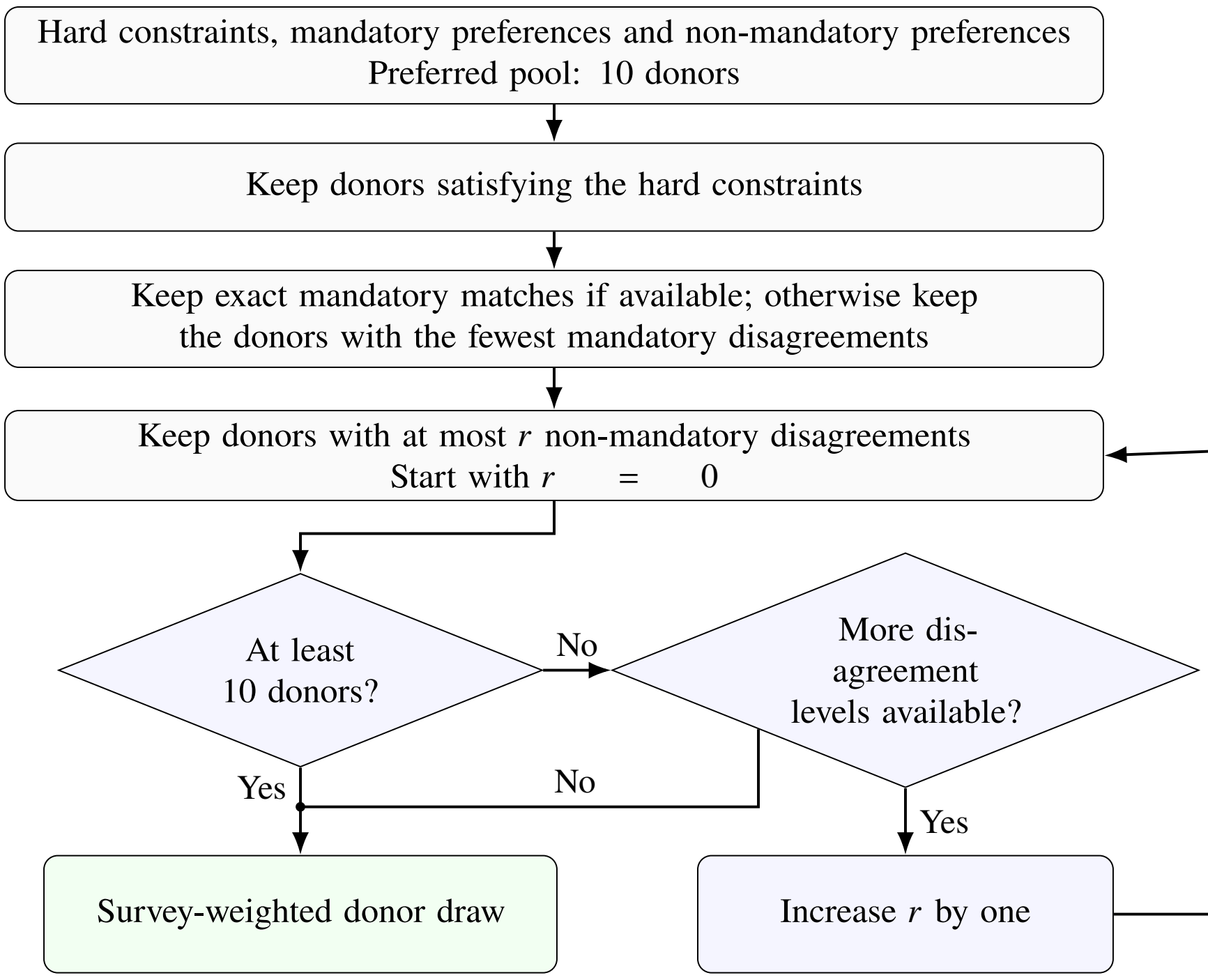

If fewer than ten donors remain at the best mandatory level, OHDM uses that smaller pool rather than accepting more mandatory disagreements merely to reach ten. It samples a donor according to survey weights. Mandatory preferences and hard constraints therefore have different roles: mandatory preferences may be relaxed when necessary, but the hard constraints remain fixed. \autoref{fig:ohdm_matching} summarises the process.

\subsection{Inputs to the learned representation}
\begin{table}[b]
\centering\small
\caption{Inputs to BSLM.}
\label{tab:features}
\begin{tabular}{p{.22\textwidth}p{.68\textwidth}}
\toprule
Input group & Variables \\
\midrule
Age and income & Exact age; original ordered household income class (0--8) \\
Household composition & Household size; adult count; numbers of children under 3, 6, 12 and 18 years \\
Density and proximity & Residents within 2.5\,km of home; usual commuting distance; distance to the Swiss border \\
Residential context & Canton; municipality type; home public-transport quality \\
Mobility resources & Number of cars in the household; car availability; driving licence \\
Work and education & Employment indicator; student indicator; employment category; work-location type; job position \\
Personal context & Sex; marital status; Swiss nationality indicator \\
\bottomrule
\end{tabular}
\end{table}

\autoref{tab:features} lists the attributes that are available for both the survey respondents and the synthetic persons, and that are used as input to the BSLM. We use exact age, the original income class, and household counts rather than broader matching bins. Residential density counts people within 2.5\,km of home. Usual commuting distance is the larger available work or education distance. Municipality type and public-transport quality describe the residential setting \citep{fso2024spatial,are2022quality}.

Work-location type distinguishes a fixed workplace, varying workplaces, home-based work and no work location. Numerical inputs are put on comparable scales, with a logarithmic transformation for density and distances. Missing values are replaced by training-sample medians and marked as missing \citep{little2019missing}. Categories use one-hot encoding: each possible category is represented by an indicator \citep{goodfellow2016deep}. These preparations allow the model to use different kinds of information together.

\subsection{Learning behavioural similarity}
The model has two parts during training. An encoder combines a person's characteristics into a short vector of 16 values, called the latent representation. A decoder uses those values to predict the person's activities and travel. Comparing these predictions with the observed diary teaches the encoder which combinations of characteristics help describe travel behaviour. People with similar predicted behaviour can then be close in this representation even if their ages, household compositions, or income classes differ.

\begin{figure*}[t]
\centering
\resizebox{.90\textwidth}{!}{\begin{tikzpicture}[>=Latex,
 input/.style={draw,rounded corners,fill=gray!6,align=center,font=\footnotesize,text=revisionred,text width=28mm,minimum height=20mm},
 block/.style={draw,rounded corners,align=center,font=\footnotesize,text=revisionred,text width=40mm,minimum height=8mm},
 training/.style={block,fill=blue!6},
 deployment/.style={block,fill=green!8}]
\node[input] (numeric) at (0,0) {Age and income\\Household size\\Numbers of children};
\node[input] (geo) at (3.2,0) {Home setting\\Density, public transport,\\municipality and canton};
\node[input] (mobility) at (6.4,0) {Mobility resources\\Number of cars, car access\\and driving licence};
\node[input] (work) at (9.6,0) {Work and education\\Employment, study,\\workplace and commute};
\node[input] (personal) at (12.8,0) {Personal characteristics\\Sex, marital status\\and nationality};
\foreach \name in {numeric,geo,mobility,work,personal}
 {\draw (\name.south) -- (\name.south |- 0,-1.2);}
\draw (0,-1.2) -- (12.8,-1.2);
\node[block] (concat) at (6.4,-1.9) {Concatenate: 48 values};
\draw[->] (6.4,-1.1) -- (concat.north);
\node[training] (dense) at (6.4,-3.1) {Dense $48\to64$\\LN, SiLU, dropout 0.2};
\node[training] (latent) at (6.4,-4.4) {Dense $64\to16$ + LN\\person representation $z$};
\draw[->] (concat) -- (dense);\draw[->] (dense) -- (latent);
\node[training] (noise) at (6.4,-5.8) {Training: $\widetilde z=z+\epsilon$\\Gaussian noise, $\sigma=0.08$};
\draw[->,blue!60!black] (latent) -- (noise);
\node[block,text width=35mm,fill=blue!3] (regularizers) at (.6,-4.5) {Training regularizers\\latent norm and\\minibatch spread};
\draw[->,dashed,blue!60!black] (noise.west) -| (regularizers.south);
\node[deployment] (retrieval) at (12.2,-4.4) {Application: normalize $z$\\employment/student mask\\200 nearest donors, $\tau=0.1$};
\node[deployment] (diary) at (12.2,-5.9) {Survey-weighted donor draw\\complete observed diary};
\draw[->,green!45!black] (latent.east) -- (retrieval.west);
\draw[->,green!45!black] (retrieval) -- (diary);
\node[training,text width=44mm,minimum height=18mm] (sequence) at (.6,-8) {Sequence branch: GRU, 16 units\\$h_0=\widetilde z$; zero inputs, $L\leq12$\\Heads: purposes, modes, departure,\\distance, travel time, dwell time};
\node[training,text width=44mm,minimum height=18mm] (count) at (6.4,-8) {Trip-count branch\\Dense $16\to8\to1$\\LN, SiLU; positive output};
\node[training,text width=44mm,minimum height=18mm] (reconstruction) at (12.2,-8) {Reconstruction branch\\Dense $16\to64\to48$\\LN, SiLU, dropout 0.2\\Detached encoder-input target};
\draw (noise.south) -- (6.4,-6.8);
\draw (.6,-6.8) -- (12.2,-6.8);
\foreach \name in {sequence,count,reconstruction}
 {\draw[->,blue!60!black] (\name.north |- 0,-6.8) -- (\name.north);}
\node[font=\footnotesize,text=revisionred,align=center] at (6.4,-9.5) {Blue: training. Green: matching uses the fitted encoder without added noise.};
\end{tikzpicture}}
\caption{BSLM network and matching process. Dense means a fully connected layer, LN is layer normalization, and SiLU is a smooth nonlinear activation. Each categorical projection also uses a dense layer, LN and SiLU. During training, the sequence and numeric output branches teach the shared representation to predict observed travel. Missing measurements are excluded from the losses. During matching, the normalized representation selects real survey diaries.}
\label{fig:workflow}
\end{figure*}
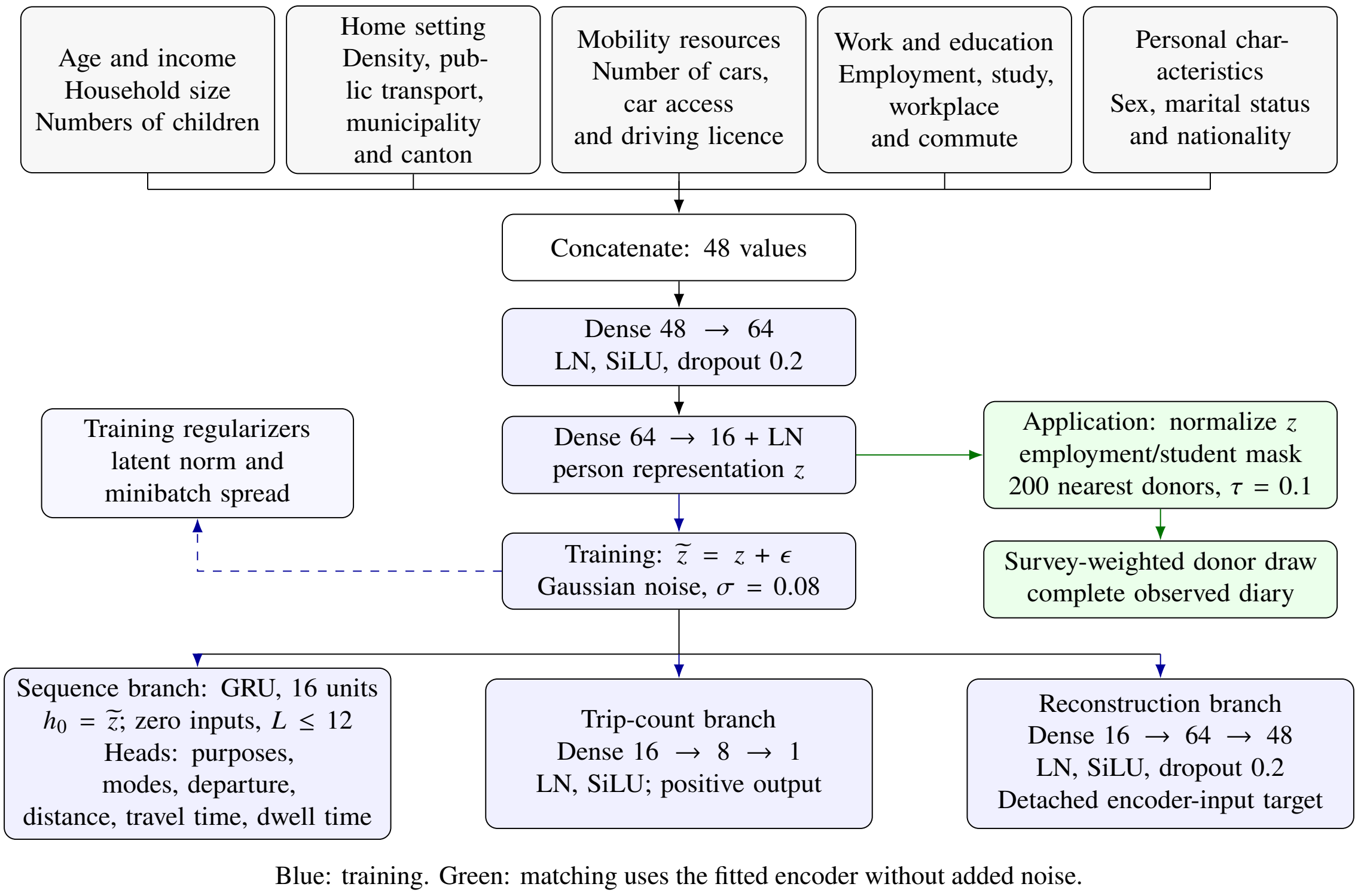

The encoder is a multilayer perceptron (MLP) that maps the input attributes of a person to a latent representation. The decoder predicts the sequence of activity purposes and modes, together with the associated departure times, trip distances, travel durations and activity durations. To carry information from one position of the sequence to the next, the decoder employs a recurrent layer based on a gated recurrent unit (GRU) \citep{cho2014gru}. An additional output head predicts the number of trips of a person. Finally, a reconstruction output is included, which encourages the latent representation to retain information about the person \citep{hinton2006reduction}. \autoref{fig:workflow} shows the main ideas; architectural and loss details are given in \ref{sec:technical}.

\subsection{Training and diary assignment}
We use 90\% of the eligible survey respondents for training and 10\% for validation, keeping households together. The model is trained once and then reused for matching. During training, prediction errors for purposes, modes, times, distances and trip counts are combined, with survey weights accounting for unequal representation. The validation sample checks how these errors develop on respondents excluded from the parameter updates. Monitoring begins after ten epochs, where an epoch is one pass through the training sample. Training stops after 12 epochs without improvement, or at 100 epochs, and retains the best validation result. In this experiment it stopped at epoch 42 and retained epoch 30.

Training uses AdamW optimizer \citep{loshchilov2019adamw}, with batches of 128 respondents, an initial learning rate of 0.005 and weight decay of 0.003. The learning rate halves every 15 epochs. Dropout, which temporarily sets some internal model values to zero during training, and small added noise reduce reliance on particular model values \citep{srivastava2014dropout,bishop1995noise}. \ref{sec:technical} records the remaining settings and equations.

To assign a diary to a synthetic person, donors and recipients are represented in the same latent space: the encoder is first applied to all survey respondents (donors) and subsequently to each synthetic person (recipient). For each recipient, the $K = 200$ nearest compatible donors in the latent space are identified, and one of them is sampled. The selection probability increases with decreasing latent distance, while the survey weights of the donors account for their representativeness in the population. The selected donor supplies the complete observed daily plan; the decoder is not used to generate a new schedule. Compatibility is enforced through a hard mask based on employment and student status. This prevents, for instance, that a non-employed person is assigned a diary containing work activities, which would cause errors in the subsequent work and education location assignment stages of eqasim. The selection probability is defined in Section~\ref{sec:technical}.

\section{Study data}
\label{sec:evaluation}
The Swiss 2015 Mobility and Transport Microcensus describes the travel of permanent residents aged six and over \citep{fso2017microcensus}. We use 37\,876 eligible respondents with domestic workday diaries. Genuine non-travel days remain in both training and evaluation. The synthetic population is a 10\% sample of the Swiss population, generated by the Swiss eqasim pipeline \citep{switzerlandScenario, dib2026onlineCalibration}. The matching comparison covers 850\,663 eligible synthetic residents.

The survey and synthetic population serve different purposes in the evaluation. A respondent has an observed diary against which alternative donor days can be scored. A synthetic resident does not. Survey-wide frequencies therefore provide a descriptive population reference, not a known correct distribution for the synthetic population. In addition, survey behaviour is observed in 2015 whereas the synthetic population represents 2025. Differences may reflect population composition, sampling and changes over time as well as the matching algorithm.

Primary and secondary purposes distinguish locations within a person's diary, not whether an activity is important. The main office is the primary work location; trips to other workplaces during the day, including business visits, are recorded as secondary work. Similarly, a primary home is distinguished from other recorded home locations, which are secondary home activities. The probability analysis includes both primary and secondary work in ``any work'', while also examining secondary work separately.

\section{Experiment 1: replaying the Microcensus}
\label{sec:replay}
\label{sec:results}
The first experiment asks whether a method can assign a suitable alternative diary to a survey respondent whose own diary is unavailable as a donor. We split the 37\,876 respondents into three groups, keeping households together. Each group receives donors only from the other two groups. Every method supplies 100 sampled diaries per respondent. Because the model has already learned from this survey, this is a replay with donor exclusion, rather than an independent test on a new survey.

\subsection{How the assigned days are assessed}
A person can have several reasonable days, so we assess the distribution of all 100 assigned diaries. The energy score compares the whole day through trip counts, purposes, modes, times and distances. The continuous ranked probability score (CRPS) assesses one quantity, such as daily distance or shopping visits, in its original units. Their equations are provided in \ref{sec:scores}. Lower values are better for every score used here \citep{gneiting2007scores}.

The Brier score checks probabilities of events such as shopping; chain Brier checks probabilities of complete activity-and-mode sequences. The variogram score additionally checks relationships between daily quantities. Their definitions and worked interpretation are collected with the other metrics in \ref{sec:scores}.

Results use survey weights and the same respondents for every method. The displayed averages divide the survey-weighted sum of individual scores by the total survey weight. The error bars are 95\% intervals obtained by resampling households 500 times \citep{davison1997bootstrap}. They describe variation across respondents for this experiment, not changes from another survey or model specification.

\subsection{Replay results}
\begin{figure}[htbp]
\centering\updatedfigure{\includegraphics[width=.8\textwidth]{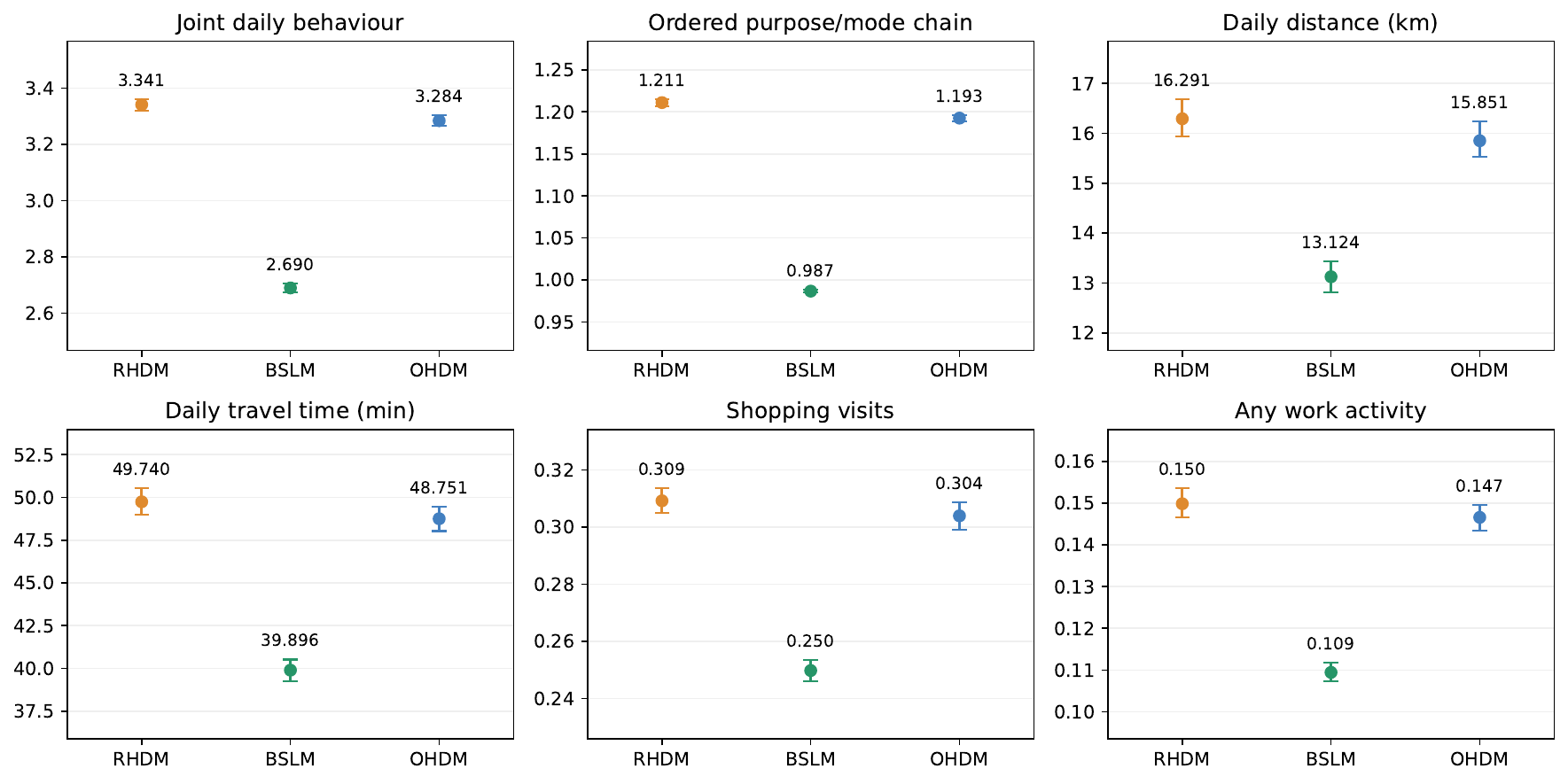}}
\caption{Microcensus replay scores for the same 37\,876 respondents. Lower is better. Each point is a survey-weighted mean over the 100-draw predictive distributions; whiskers are 95\% household-bootstrap intervals.}
\label{fig:replay}
\end{figure}

BSLM has lower replay errors than both categorical methods across the six main outcomes in Figure~\ref{fig:replay}. Its energy score is 2.690, compared with 3.341 for RHDM, a reduction of \EnergyGain\%. Chain Brier decreases from 1.211 to 0.987 (\ChainGain\%), and daily-distance CRPS decreases from 16.291 to 13.124\,km (\DistanceGain\%). Travel-time CRPS falls by 19.8\%, shopping-count CRPS by 19.2\%, and work-event Brier by 27.0\%. These are improvements in reconstruction under donor exclusion.

OHDM provides smaller improvements over RHDM. Its energy score is 3.284, chain Brier 1.193 and distance CRPS 15.851\,km, corresponding to reductions of approximately 1.7\%, 1.5\% and 2.7\%. Removing a fixed relaxation order therefore changes the results, but does not close the replay-score gap to BSLM in this experiment. We assess uncertainty using a paired household bootstrap with 500 resamples. In each resample, households are drawn with replacement, keeping their members together, and the survey-weighted mean score difference between BSLM and RHDM is recalculated for the same respondents. The central 95\% of these estimates gives intervals of $[-0.664,-0.638]$ for energy, $[-0.228,-0.221]$ for chain Brier and $[-3.345,-3.009]$\,km for distance CRPS. All three intervals lie below zero, indicating lower prediction errors for BSLM.

\subsection{Agreement between predicted probabilities and observed frequencies}

This diagnostic assesses whether the activity probabilities implied by donor matching agree with the frequencies observed in the survey diaries. For each respondent $i$ and activity type $a$, the predicted probability $\hat{p}_{ia}$ is the share of the $K = 100$ assigned donor days that contain activity $a$. For example, if 30 of the 100 donor days contain a shopping activity, the predicted shopping probability is $\hat{p}_{i,\mathrm{shop}} = 30/100 = 0.3$. The observed outcome $y_{ia} \in \{0,1\}$ indicates whether activity $a$ appears as the origin or destination of at least one trip in the respondent's recorded diary day.

For all methods, respondents are initially grouped into the same probability bins $[0, 0.1)$, $[0.1, 0.2)$, $[0.2, 0.3)$, $[0.3, 0.5)$, $[0.5, 0.75)$ and $[0.75, 1]$. Adjacent bins are then merged until each resulting bin contains at least 20 respondents for every method. Merging uses predicted-probability counts, not observed outcomes. The resulting boundaries remain common to all three methods within each activity panel, but may differ between activity types. For a bin $b$ with respondent set $\mathcal{I}_b$ and survey weights $w_i$, the mean predicted probability and the observed frequency are given by
\begin{equation}
\bar{p}_{ab} = \frac{\sum_{i \in \mathcal{I}_b} w_i \, \hat{p}_{ia}}{\sum_{i \in \mathcal{I}_b} w_i}, \qquad
\bar{y}_{ab} = \frac{\sum_{i \in \mathcal{I}_b} w_i \, y_{ia}}{\sum_{i \in \mathcal{I}_b} w_i}.
\end{equation}

\begin{figure}[!h]
\centering
\updatedfigure{\includegraphics[width=.8\textwidth]{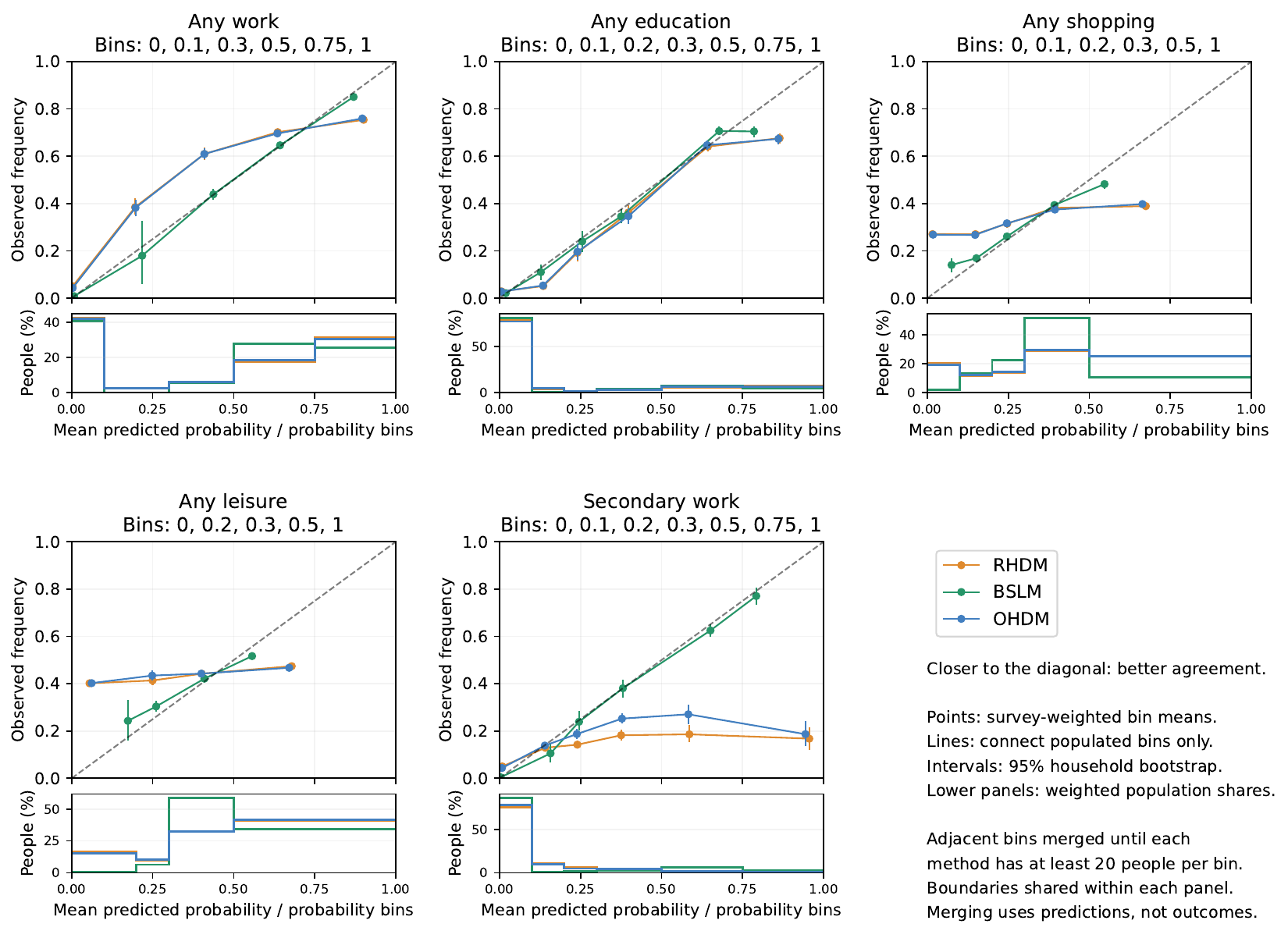}}
\caption{Agreement between predicted probabilities and observed frequencies for five diary events. The upper panels show the mean predicted probability against the observed frequency for common probability bins. Vertical bars indicate pointwise 95\% household-bootstrap intervals, conditional on the fitted encoder, the donor draws and the bin membership; they are not adjusted for multiple comparisons. The lower panels show the survey-weighted share of respondents in each bin, expressed as percentages rather than probability densities, such that the bars sum to 100\% for each method. Each plotted bin contains at least 20 respondents per method after adjacent-bin merging.}
\label{fig:probability_agreement}
\end{figure}

In the upper panels of \autoref{fig:probability_agreement}, each point is located at $(\bar{p}_{ab}, \bar{y}_{ab})$. Points below the diagonal indicate overestimated probabilities, whereas points above it indicate underestimated probabilities. The mean predicted probability is used instead of the bin midpoint, such that the diagonal serves as a direct reference for agreement \citep{gneiting2007calibration}.

For BSLM, respondents in the highest shopping-probability bin, $[0.5,1]$, have a mean predicted probability of 54.7\%, whereas the observed shopping frequency is 48.2\% ($n = 4\,146$). For leisure, the same probability range has a mean prediction of 55.7\% and an observed frequency of 51.7\% ($n = 13\,302$). The small extreme-probability groups are included in these wider bins rather than displayed as separate points. Conversely, respondents in the low-probability bins tend to exhibit higher observed frequencies than predicted. A low replay error is therefore not sufficient to ensure agreement at all probability levels. The histograms indicate how many respondents contribute to each part of the curves, and the intervals quantify the uncertainty of the observed frequencies.

\subsection{How clear are the differences between methods?}
\begin{figure}[htbp]
\centering\updatedfigure{\includegraphics[width=.96\textwidth]{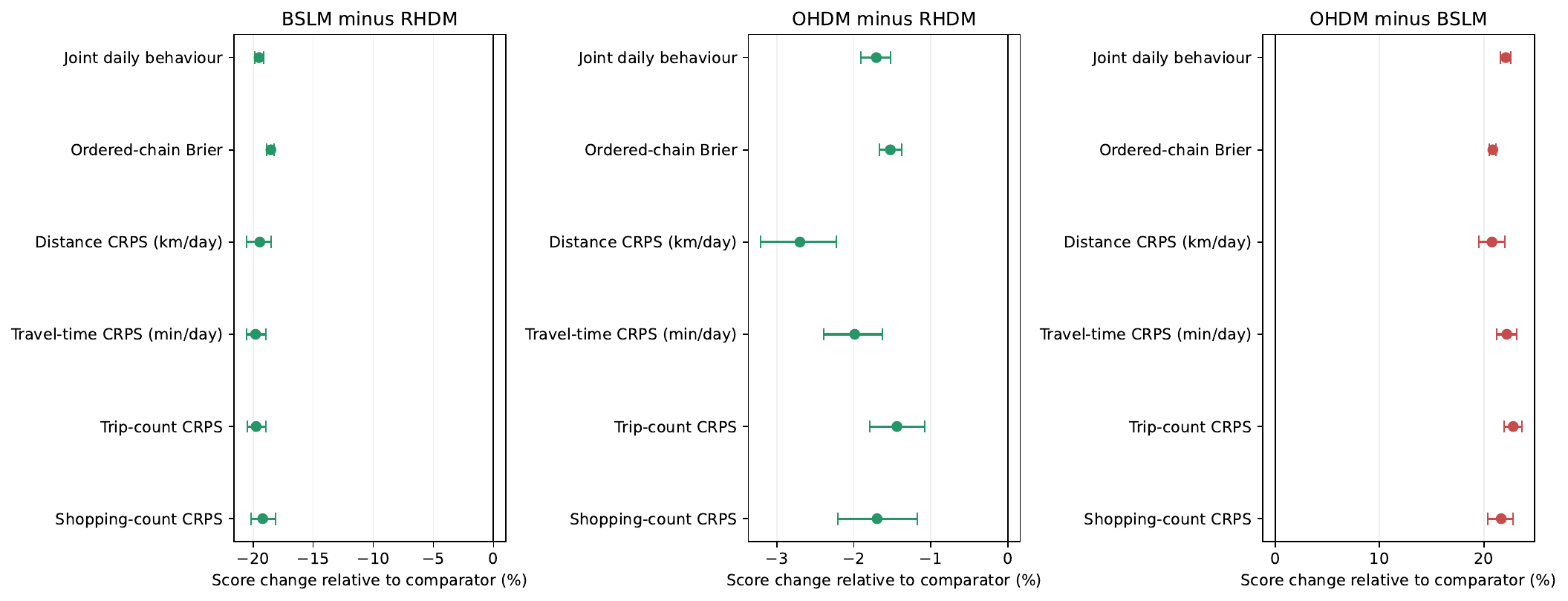}}
\caption{Paired score differences, expressed relative to the comparator. Green intervals are below zero, red intervals are above zero. Further left favours the first named method.}
\label{fig:paired}
\end{figure}

Figure~\ref{fig:paired} compares the errors of the methods for the same respondents. Negative values favour the first-named method, since lower scores indicate better performance. BSLM achieves lower errors than the other methods across all metrics shown in the figure. OHDM also outperforms RHDM, although the improvement is less pronounced than for BSLM.

\subsection{Does replay performance vary by canton?}
\begin{figure}[htbp]
\centering\updatedfigure{\includegraphics[width=.9\textwidth,keepaspectratio]{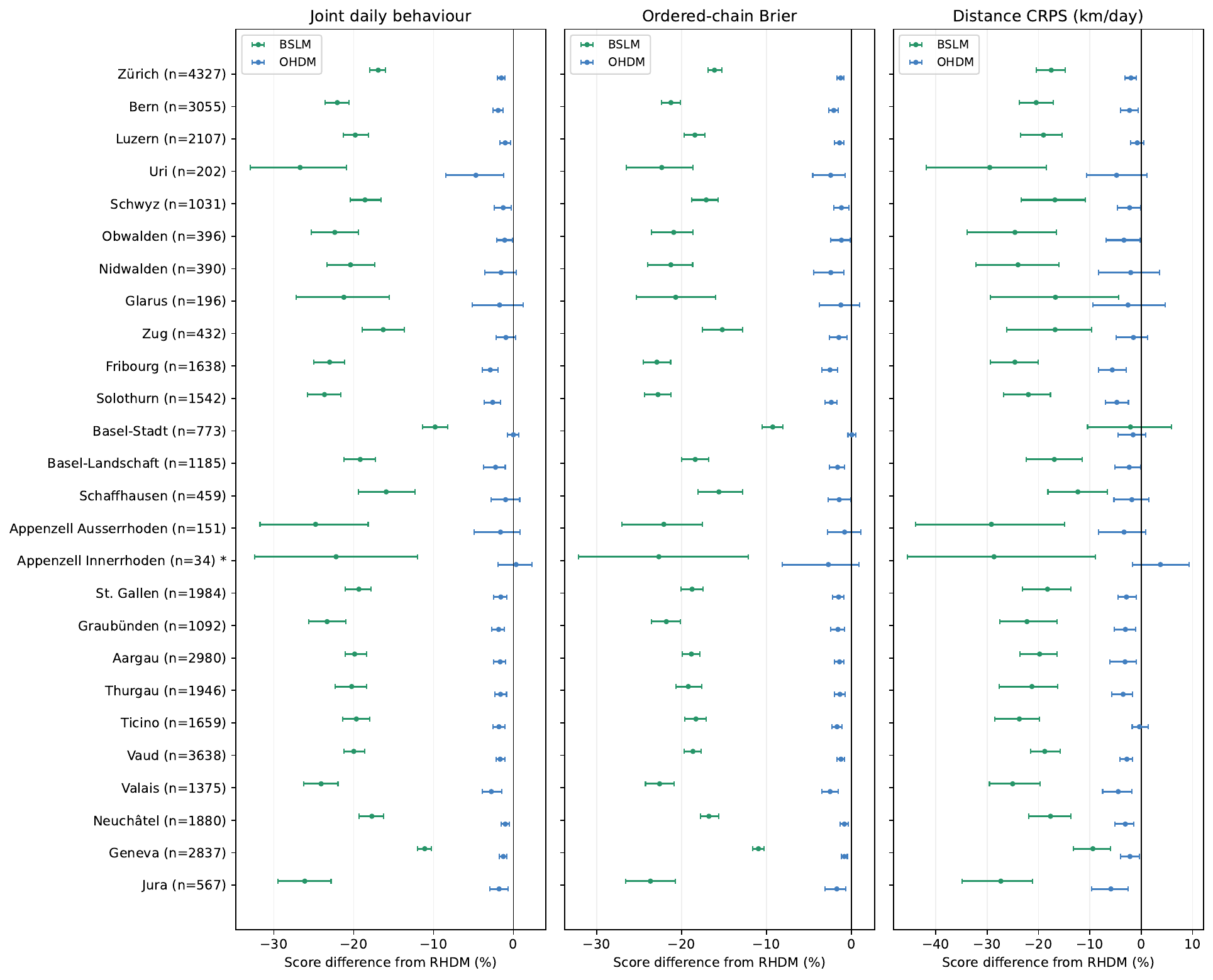}}
\caption{Canton-level replay score differences from RHDM. Negative values favour BSLM or OHDM. Whiskers are 95\% household-bootstrap intervals; $n$ is the number of survey respondents. An asterisk marks a canton with fewer than 100 respondents.}
\label{fig:canton_scores}
\end{figure}

Figure~\ref{fig:canton_scores} repeats the comparison within each respondent's home canton. It shows changes from RHDM for joint behaviour, complete chains and daily distance, with household-bootstrap intervals and respondent counts. This distinguishes the size of the estimated difference from its uncertainty. Cantons with small survey samples should not be ranked on their point estimates alone.
BSLM has lower point estimates than RHDM in all 26 cantons for the three displayed scores. The intervals lie below zero in all 26 cantons for joint behaviour and ordered chains, and in 25 for daily distance. 

\section{Experiment 2: matching the synthetic population}
\label{sec:population}
We next ask what the matching methods imply for the synthetic population. Each method assigns diaries to the same 850\,663 eligible residents in the configured 10\% population sample, repeating the assignment twenty times. We first compare Switzerland as a whole, then Geneva, Z\"urich and Vaud cantons. The repetitions show variation caused by random donor selection.

We compare trips per day, total daily distance, visits by purpose and complete activity chains. Each arrival at an activity counts as a visit, including repeat visits. Distance is the sum of straight-line trip distances, including returns home, not distance travelled on a routed network. The survey provides a descriptive reference because the true diaries of synthetic residents are unknown. Figure ranges show the minimum and maximum across twenty assignments.

\subsection{National and canton results}
The twenty population repetitions produce very similar national trip counts: 3.650 trips per person-day with RHDM, 3.653 with BSLM and 3.633 with OHDM. The BSLM--RHDM difference is only 0.08\%. Daily straight-line distance differs more: 22.283, 23.078 and 22.212\,km, respectively. BSLM therefore adds 3.57\% relative to RHDM despite the near-equal trip count. The survey reference is 3.654 trips and 23.091\,km. Agreement on the number of trips does not establish agreement on how far those trips extend.

\begin{figure}[htbp]
\centering\updatedfigure{\includegraphics[width=\textwidth]{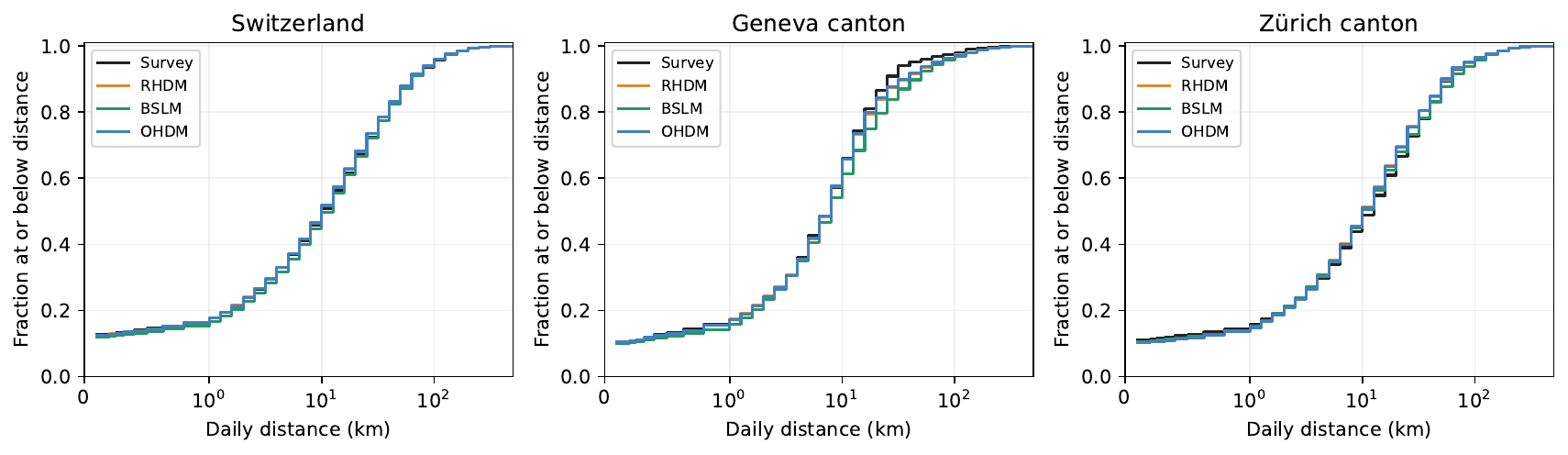}}
\caption{Daily straight-line-distance distributions nationally and by home canton. Curves are cumulative histogram shares, averaged over matching repetitions. Zero-distance days are included. Closer to the black survey curve indicates descriptive distribution agreement, not a validated population forecast.}
\label{fig:distances}
\end{figure}

Figure~\ref{fig:distances} compares the daily-distance distributions behind these means. The reference and matched curves are often close at national scale. These aggregate curves complement the replay distance scores: a plausible marginal distance distribution does not establish appropriate distance assignments to particular people.

\begin{figure}[htbp]
\centering\updatedfigure{\includegraphics[width=.90\textwidth]{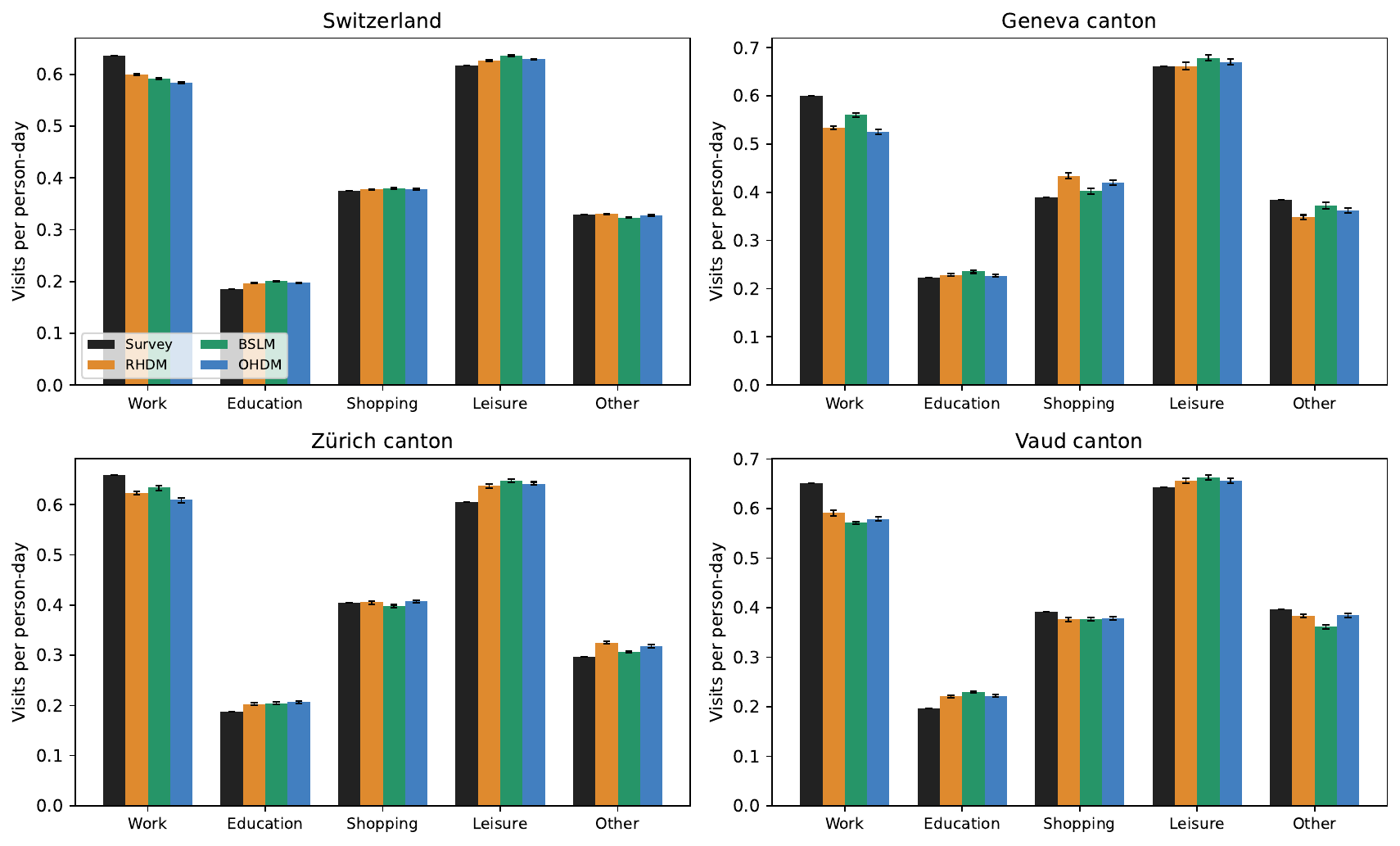}}
\caption{Mean visits per person-day nationally and by home canton. Black bars show the survey reference; coloured bars average twenty synthetic-population matching repetitions. Whiskers span the minimum and maximum over repetitions. Closer to black means aggregate survey agreement only, not proven population accuracy. Zero-visit residents remain in every mean.}
\label{fig:population}
\end{figure}

Figure~\ref{fig:population} shows similarly close national purpose totals, but different regional changes. National work visits decrease from 0.599 with RHDM to 0.592 with BSLM (1.24\% lower), while shopping visits increase from 0.378 to 0.380 (0.65\% higher). In Geneva canton, however, work visits increase from 0.534 to 0.561, and shopping visits decrease from 0.433 to 0.403. Thus, the direction of change is opposite to the national result for both purposes. In Vaud, work visits decrease from 0.592 to 0.571. In Z\"urich canton they change only slightly, from 0.623 to 0.634. The same national method comparison can therefore have different implications for local demand composition.

\begin{figure}[htbp]
\centering\updatedfigure{\includegraphics[width=.85\textwidth]{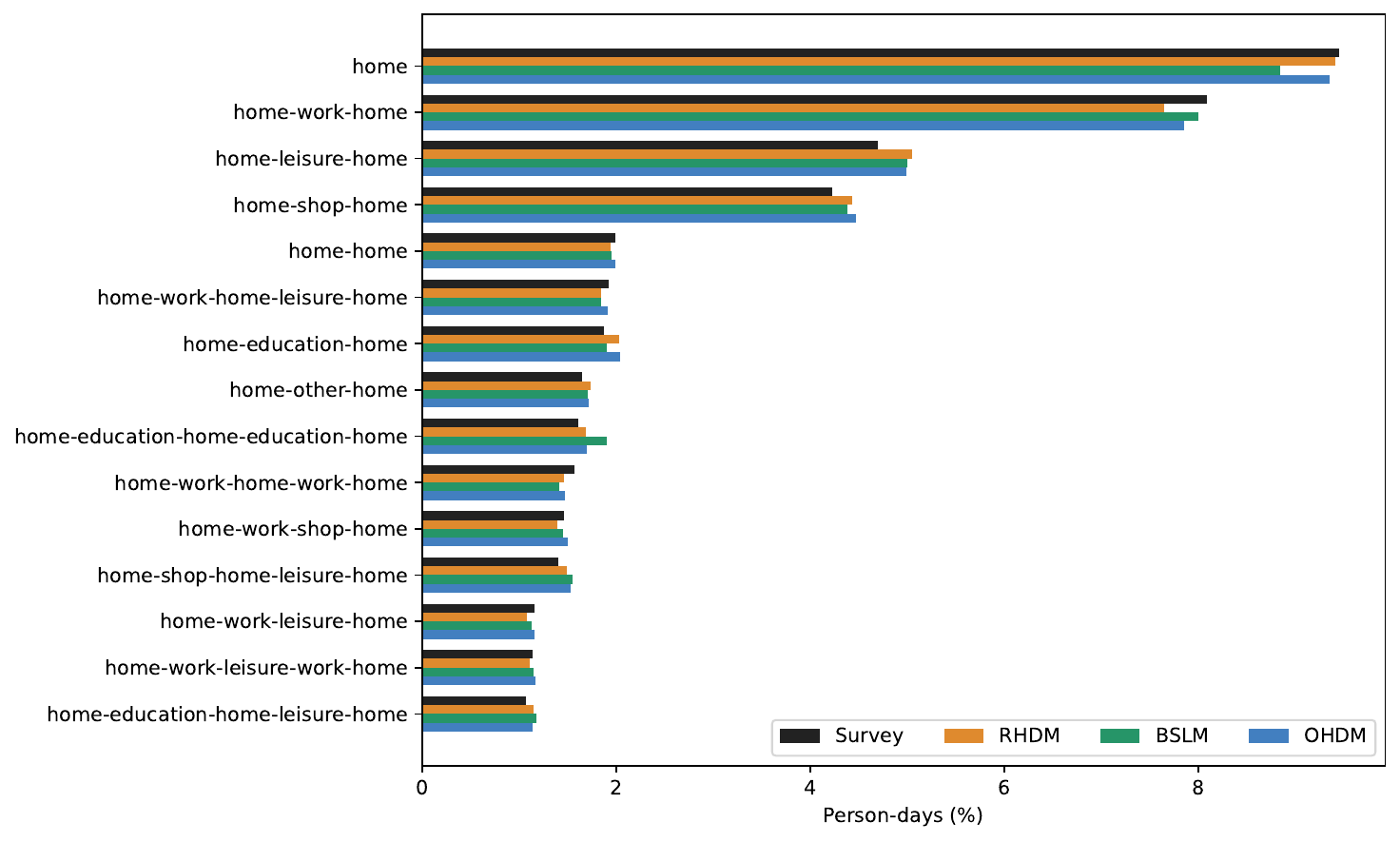}}
\caption{The fifteen most frequent survey purpose chains, compared with synthetic-population shares averaged across twenty repetitions.}
\label{fig:chains}
\end{figure}

The chain comparison in Figure~\ref{fig:chains} gives another aggregate view, but does not provide a universal ranking of matching methods. For example, the survey work-visit mean is 0.636: all three population means are lower, and RHDM is closer to that particular reference than BSLM. This does not negate BSLM's lower replay errors. The measures ask different questions: how closely population totals resemble the survey, and how well another donor's day represents a particular respondent. Neither question alone establishes true population behaviour.

\section{Experiment 3: examining local differences}
\label{sec:local}
The third experiment examines whether the national comparison describes smaller areas. We group residents within 2\,km of Z\"urich HB and Gen\`eve-Cornavin. These central areas contain 9\,643 and 13\,899 eligible synthetic residents in the 10\% population sample, respectively; these are sample counts, not the full populations of the central areas. We also group homes within 4\,km of each station into 1\,km cells, retaining cells with at least 100 residents and an RHDM mean of at least 0.05 visits per person-day. This gives 44 cells around Z\"urich and 37 around Geneva. 

These comparisons describe the complete diaries of people who live in each area. They do not count activities or traffic inside the area. Positive differences mean more distance or visits, not better performance. Ranges across repetitions describe random matching variation.

\subsection{Central-area and cell-level results}
\begin{figure}[htbp]
\centering\newrevisionfigure{\includegraphics[width=0.9\textwidth]{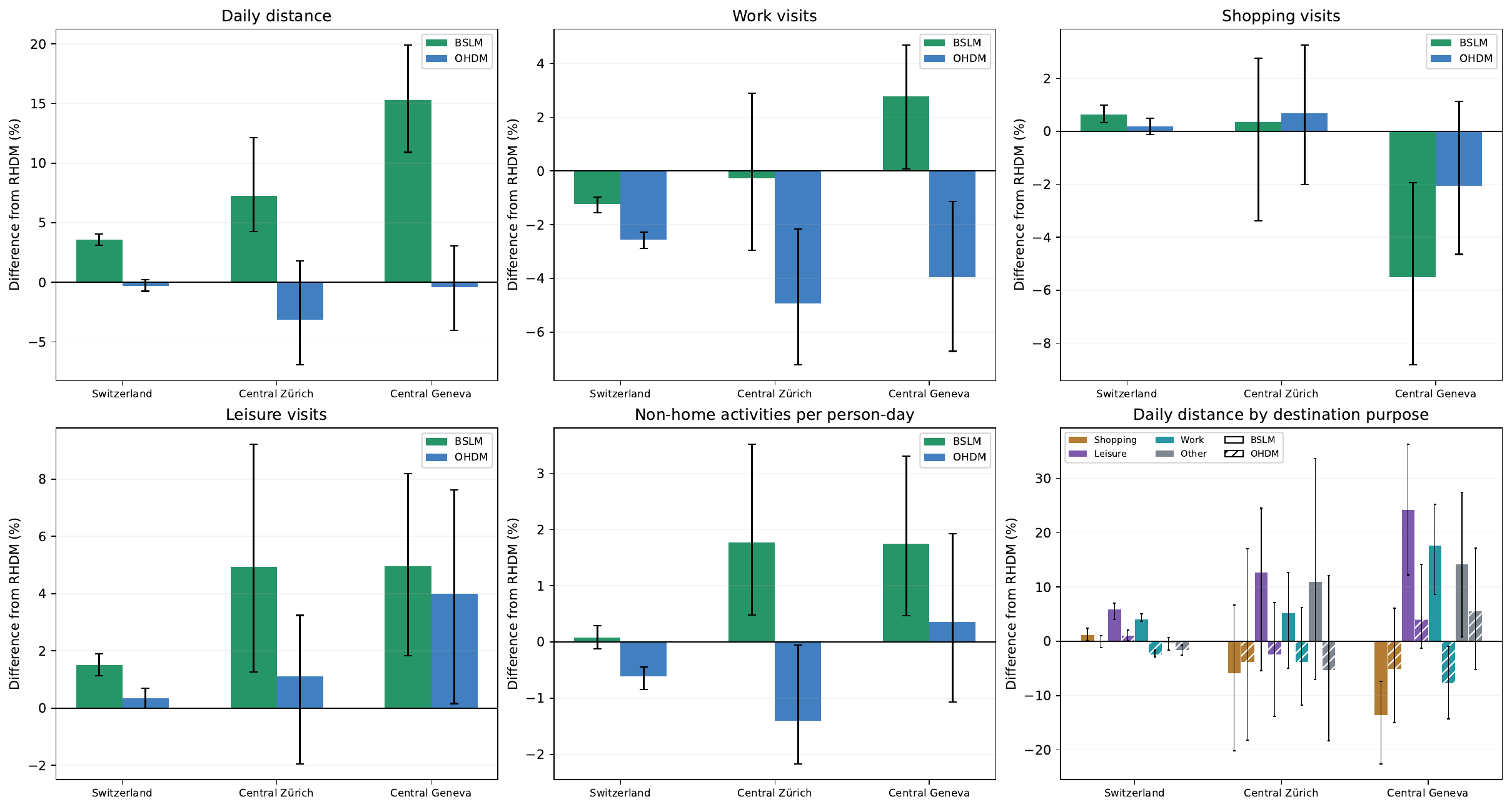}}
\caption{\newrevision{Differences from RHDM for national and central-area residents. Bars show mean relative differences over twenty matching repetitions; whiskers show their range. Activities count non-home destination visits per person-day, including repeated visits. Distance is straight-line travel to the selected destination purposes. In the purpose-distance panel, colours distinguish purposes and hatching identifies OHDM. Central areas cover residents within 2\,km of the two stations in the configured 10\% population sample. Outcomes describe residents' complete diaries, not activities inside the circles.}}
\label{fig:local}
\end{figure}

Figure~\ref{fig:local} moves from cantons to residents within 2\,km of the two central stations. In central Geneva, BSLM produces 5.52\% fewer shopping visits and 4.94\% more leisure visits than RHDM. Work visits increase by 2.77\%, reversing the national decrease. Mean daily distance rises from 13.402 to 15.446\,km, a 15.25\% increase. In central Z\"urich, shopping visits instead increase by 0.33\%, leisure visits by 4.93\%, and education visits by 11.90\%. Daily trip counts rise by 0.65\% in central Geneva and 1.10\% in central Zürich, compared with 0.08\% nationally.

\begin{figure}[htbp]
\centering\updatedfigure{\includegraphics[width=.85\textwidth]{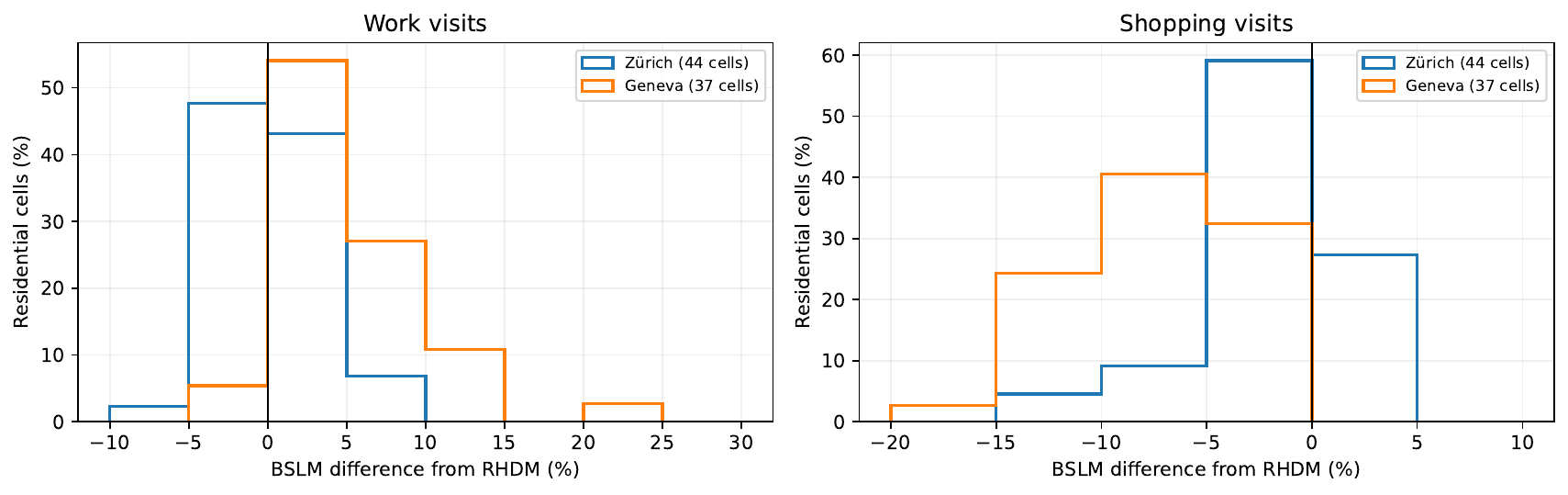}}
\caption{Distribution of BSLM--RHDM differences across residential 1\,km cells around each station. Each cell has equal weight, at least 100 eligible residents, and an RHDM mean of at least 0.05 visits per resident-day. Histograms show the percentage of cells in each 5-percentage-point bin, using differences between the twenty-repetition mean counts. The bin percentages sum to 100\% for each city. Positive and negative values describe directional sensitivity, not accuracy. }
\label{fig:cells}
\end{figure}

The cell-level comparison in Figure~\ref{fig:cells} shows further variation within these urban settings. For Geneva shopping visits, the middle 80\% of cell-level differences ranges from $-12.5\%$ to $-3.0\%$, with a median of $-7.0\%$. In Z\"urich it ranges from $-6.2\%$ to $2.7\%$, with a median of $-1.8\%$. Individual eligible cells range from $-17.3\%$ to $-1.3\%$ in Geneva and from $-11.6\%$ to $4.4\%$ in Z\"urich. These ranges are spatial variation, not confidence intervals. They show why a national shopping increase of 0.65\% does not describe the direction or size of every local change.

\FloatBarrier
\subsection{\newrevision{Sensitivity to the random seed}}
\label{sec:seed-sensitivity}
\begin{figure}[htbp]
\centering\newrevisionfigure{\includegraphics[width=\textwidth]{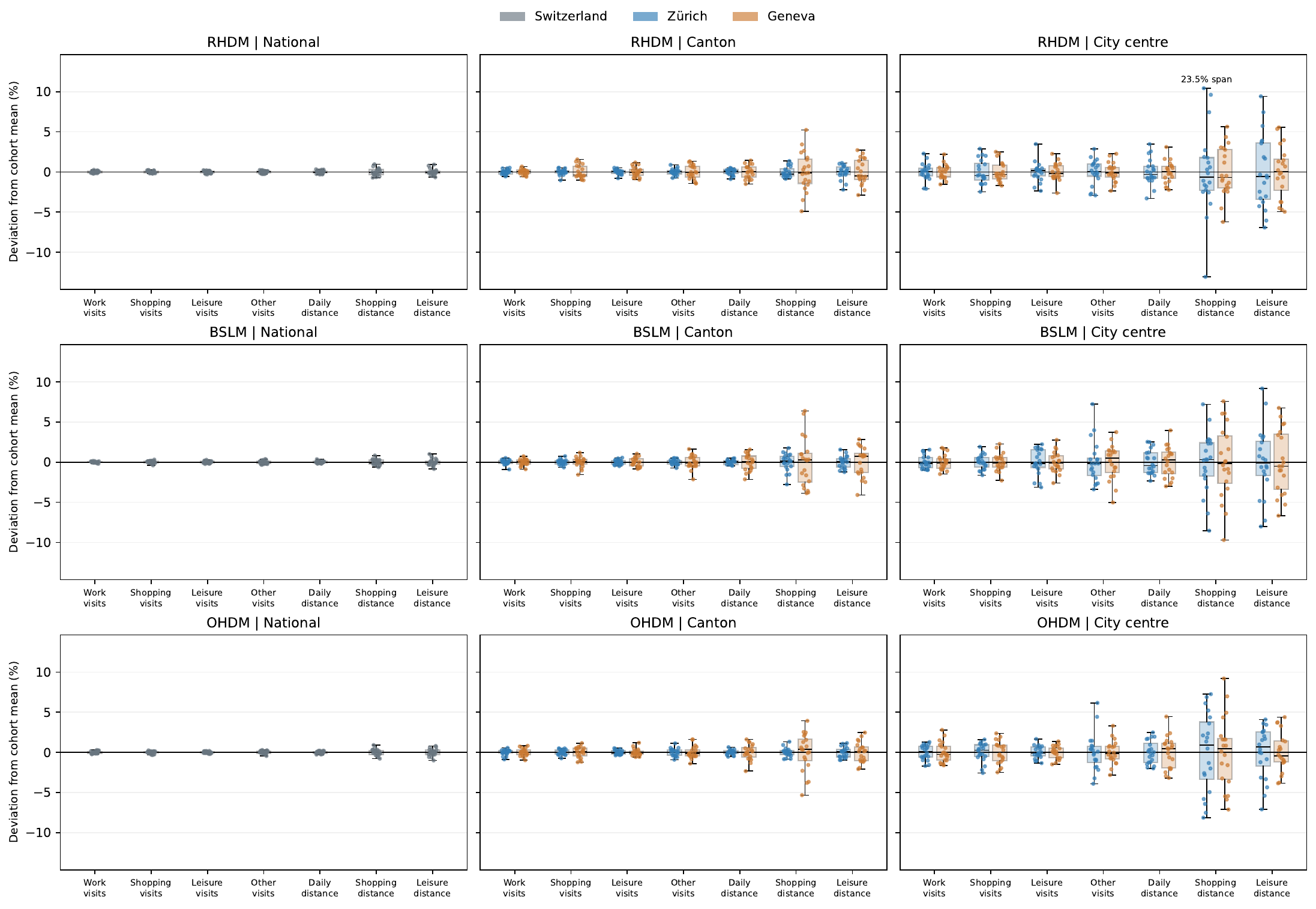}}
\caption{\newrevision{Random-seed sensitivity across twenty assignments. Rows show matching methods; columns show Switzerland, cantons and city centres. Colours identify Switzerland, Z\"urich and Geneva. Points are individual runs, relative to each method and cohort's own mean. Boxes show medians and interquartile ranges; whiskers mark the observed minimum and maximum. Shopping and leisure distance are daily straight-line kilometres to those destinations. All panels share a vertical scale. The fixed sample contains 850,663 eligible residents nationally; 152,833 and 48,852 in the two cantons; and 9,643 and 13,899 in the city centres.}}
\label{fig:seed-sensitivity}
\end{figure}

\newrevision{We repeat each population assignment twenty times with different random seeds, keeping the population, donor data and fitted encoder fixed. Figure~\ref{fig:seed-sensitivity} compares the resulting variation at three spatial scales: Switzerland, the cantons of Z\"urich and Geneva, and residents of their two city centres. For each method and geographic cohort, the twenty runs are expressed as percentage deviations from that cohort's own mean. The outcomes are work, shopping, leisure and other visits, total daily distance, and daily distance to shopping and leisure destinations. All panels share the same vertical scale. The coefficient of variation (CV), defined as the standard deviation across runs divided by their mean, provides a summary of relative variation.}

\newrevision{Across the three methods and seven outcomes, CVs range from 0.088--0.447\% nationally, 0.274--3.039\% at canton level, and 0.787--5.220\% in the city centres. National averages therefore vary little across seeds, while differences between individual runs become more visible in smaller resident groups. This pattern concerns donor-sampling variation at the fixed population sample used here; it does not measure the effects of changing the population sample size or regenerating the population with different random seeds. These additional sources could produce larger local variation, but neither is tested here.}

\newrevision{The observed minimum and maximum make the size of this variation more tangible. In central Geneva, daily distance ranges from 13.103 to 13.819\,km per person-day for RHDM, 14.982 to 16.058\,km for BSLM, and 12.925 to 13.945\,km for OHDM. The minimum-to-maximum spans represent 5.340\%, 6.966\% and 7.642\% of each method's twenty-run mean, respectively. In central Z\"urich, the corresponding spans are 6.760\%, 4.862\% and 4.485\%. For example, the OHDM Geneva points extend from $-3.185\%$ to $+4.457\%$ in Figure~\ref{fig:seed-sensitivity}, a span of 7.642 percentage points.}

\begin{figure}[htbp]
\centering\newrevisionfigure{\includegraphics[width=0.85\textwidth]{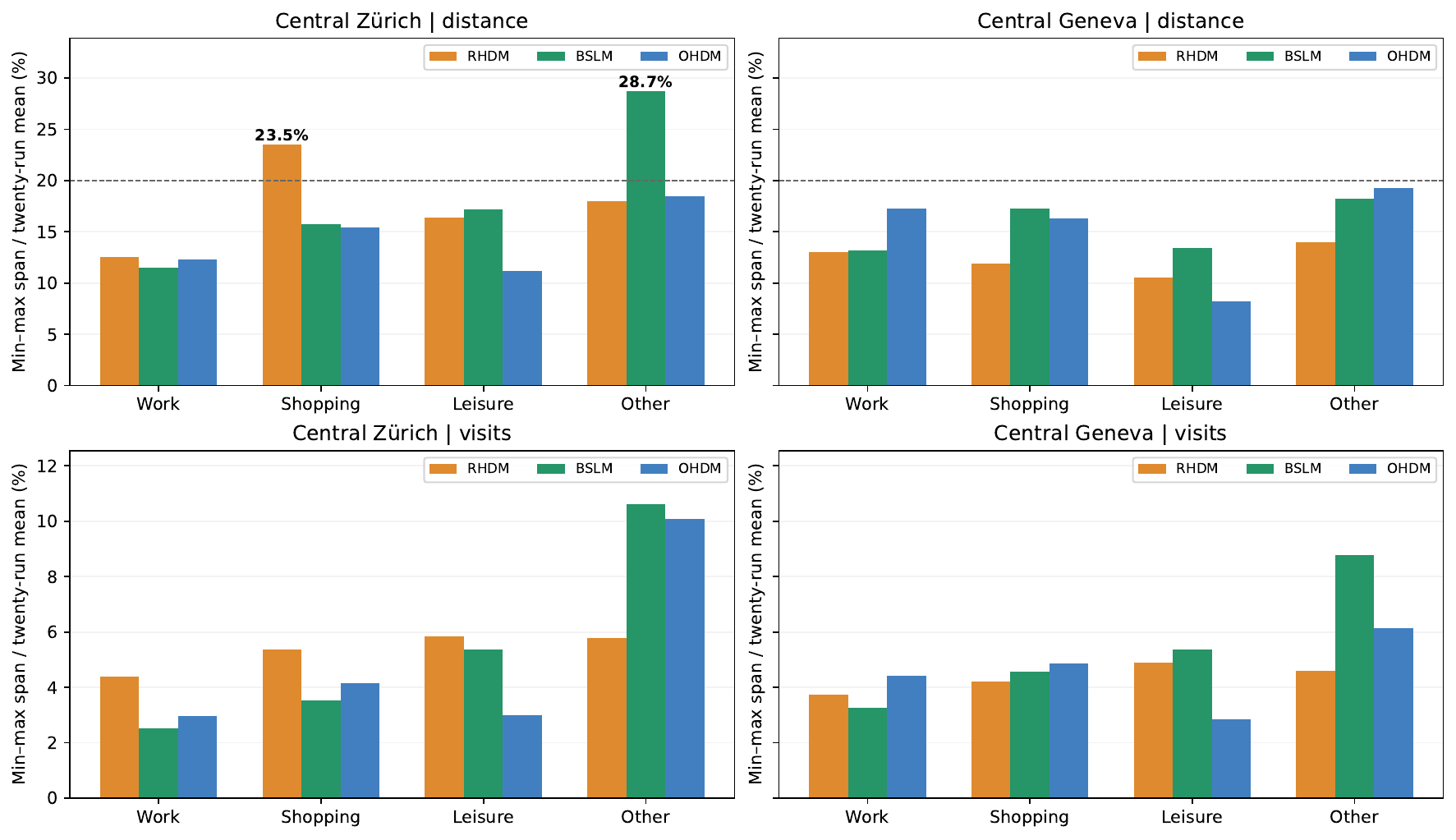}}
\caption{\newrevision{Purpose-specific variation across twenty diary assignments for city-centre residents. Bars show the observed maximum minus minimum, divided by the method's twenty-run mean. Upper panels show daily straight-line distance by destination purpose; lower panels show visit frequencies. The dashed line marks a 20\% span. Vertical scales differ between rows. }}
\label{fig:purpose-seed-spans}
\end{figure}

\newrevision{Purpose-specific outcomes show a larger spread than total daily distance (Figure~\ref{fig:purpose-seed-spans}). For BSLM in central Z\"urich, daily distance to other destinations ranges from 0.871 to 1.159\,km per person-day, a span of 28.703\% of the twenty-run mean. For RHDM in central Z\"urich, daily distance to shopping destinations ranges from 1.251 to 1.589\,km per person-day. This span is 23.498\% of the twenty-run mean; the highest run is 27.029\% above the lowest. The shopping-distance spans for BSLM and OHDM in the same area are 15.748\% and 15.389\% of their respective means. Other sizeable spans include BSLM shopping distance in central Geneva (17.283\%) and BSLM leisure distance in central Z\"urich (17.198\%). Visit frequencies vary less across the purpose groups shown: the largest span is 10.627\% for BSLM other visits in central Z\"urich. Thus, fairly stable activity counts can coexist with appreciably different distances to those activities. These are observed extrema among twenty assignments, not confidence intervals or estimates of population-sampling uncertainty.}

\section{Discussion}
\label{sec:discussion}
\subsection{What the replay establishes}
The replay results support BSLM as a useful way to retrieve alternative observed diaries. It gives lower joint, chain, distance, time and activity-specific errors than the categorical comparators in the present experiment. The result is not simply closer national totals: replay scores evaluate each recipient's distribution of donor days against their own observed day.

Two limitations apply. First, the replay assesses alternative donors after the model has been trained on the same survey. It is therefore a reconstruction test and not an independent estimate of out-of-sample accuracy. Applying BSLM to a different survey wave or region is the appropriate next test of transferability.

Second, BSLM differs from the categorical procedures in two respects other than the neural representation: the compatibility rules for donors and the definition of candidate similarity. The performance gap can therefore not be attributed to the learned representation alone. A controlled comparison with identical donor restrictions, alternative representation models and several training seeds is needed to isolate the contribution of each design choice. The OHDM comparator is informative in this respect. Relaxing the categorical matching rules alone improves replay scores modestly and allows fast population matching without representation training. Part of the  gain observed for BSLM may thus come from the matching design rather than from learned representations.

\subsection{Global agreement and local planning inputs}
The population application provides a complementary conclusion. National trip totals are nearly identical for RHDM and BSLM, but differences in distance and purpose composition become clearer when residents are grouped geographically. Central Geneva and central Z\"urich show different shopping responses, while residential cells display a wider range than either national or central-area means. Aggregation can therefore conceal both the direction and magnitude of sensitivity to diary assignment.

This matters because transport planning is often local even when models cover a whole country. Work-related demand contributes to the potential market for commuter services; shopping and leisure diaries influence the timing and composition of demand for other services and access policies. A model that reproduces a national trip count can still supply different demand inputs for neighbourhoods served by a particular intervention. The results justify treating diary matching as part of model sensitivity analysis, rather than only as a technical preprocessing choice.

The evidence does not show that the larger local differences are closer to reality. Synthetic residents can legitimately have a different demographic composition from survey respondents. Survey weights improve representation but do not guarantee unbiased behaviour for every neighbourhood, household type or time period. Conversely, a deviation should not be accepted merely because the survey may be imperfect. The practical response is to combine replay diagnostics with local empirical evidence, such as purpose-specific survey estimates where sample sizes permit, observed travel counts, travel times and relevant service-use data.

\subsection{Observed-day transfer and downstream modelling}
Matching retains combinations of activity purposes, modes and timing that were observed together. It also keeps the demand-synthesis structure compatible with eqasim. These are reasons to retain whole-day transfer while improving donor selection. They are not a guarantee that every copied day is suitable for every recipient. Usual locations, household coordination and future behavioural change remain additional modelling questions. A fixed donor library also limits the range of behaviours that can be represented.

Our local figures concern the diaries generated by residents of an area. Those residents may work, shop or spend leisure time elsewhere. Later location assignment, mode choice and MATSim simulation can redistribute or modify the resulting demand \citep{horl2023locations,horni2016matsim}. We have not measured changes in central-area arrivals, link flows, congestion, accessibility or the benefits of a particular policy. Those require further downstream scenario comparisons. The present result is that the inputs to such comparisons are sensitive to matching at a spatial scale relevant to planning.

\subsection{\newrevision{Aggregation, random seeds and population sampling}}
\newrevision{The seed comparison in Figure~\ref{fig:seed-sensitivity} shows why stable aggregate results are not sufficient evidence of stable local behaviour. With the same population and matching method, national means for work, shopping, leisure and other visits, and for distance, vary little across twenty assignments, whereas canton and city-centre averages show greater variation. Aggregation can conceal changes that matter for a smaller group of residents. The purpose-specific comparison in Figure~\ref{fig:purpose-seed-spans} further shows that a stable number of visits need not imply stable travel distances: the BSLM distance span for other destinations reaches 28.703\% of its mean in central Z\"urich. A single run that appears representative nationally may therefore give a different picture of the demand relevant to a local intervention.}

\newrevision{Agent-based transport simulations commonly use population subsamples to reduce runtime. \citet{llorca2019scaling} show that population scaling in MATSim can affect travel times and link volumes, alongside runtime. This distinction therefore also matters when selecting population sample sizes for agent-based transport models. Representing individuals explicitly does not by itself establish that one sampled population represents every behaviour the model is intended to describe. A sample may reproduce total trips while giving a less reliable account of particular activity purposes, population groups, locations or times of day. Expanding its totals to the full population does not establish that these behavioural patterns are representative. Stability across random seeds and agreement with observations should therefore be assessed at the scale and for the outcomes relevant to the intended application.}

\newrevision{Our experiment changes only the diary-matching seed, using the same fixed 10\% synthetic-population sample in every run. It does not vary the population-regeneration seed or sample size, or test the resulting traffic simulations. Regenerating the population with different seeds, or changing the sampling fraction, could produce larger variation and compound local differences. This is a further source of uncertainty, motivated by the scaling effects reported by \citet{llorca2019scaling}, rather than an effect measured in our experiment. Establishing an appropriate sample size would require those comparisons, with repeated seeds and checks on both aggregate and disaggregated outcomes. The present results motivate that assessment; they do not identify a minimum representative sample size or imply that a larger sample alone resolves behavioural differences between matching methods.}

\subsection{Uncertainty and transfer to other settings}
The twenty population repetitions describe stochastic donor selection with a fixed population, donor library and encoder. \newrevision{Their ranges do not capture uncertainty from population regeneration or changes in sample size, survey selection, training or future behaviour.} Similarly, the replay bootstrap intervals condition on the chosen encoder and donor partition. Their small width should not be read as certainty about model transfer. 

The method is applicable where both population and survey records share informative personal, household and residential attributes. Its usefulness elsewhere will depend on the donor sample, available predictors and the purpose of the transport model. The transport-planning lesson is broader than this particular network architecture: national agreement should be assessed alongside individual matching diagnostics, spatial sensitivity and evidence relevant to the decisions for which demand forecasts will be used.

\section{Conclusion}
\label{sec:conclusion}
Assigning or generating daily plans is a central step in agent-based and activity-based transport modelling. It determines how synthetic residents become travel demand. This study proposes BSLM, which learns a 16-dimensional behavioural representation and uses it to select complete observed survey diaries. It retains the structure of the eqasim pipeline while replacing categorical similarity with similarity informed by multiple travel outcomes.

In the Swiss Microcensus replay, BSLM reduces energy, chain and distance scores by \EnergyGain\%, \ChainGain\% and \DistanceGain\% relative to the default RHDM reference. These replay results assess matching to alternative survey diaries; they do not establish accuracy for a new population or future year. OHDM provides a simpler categorical comparator with smaller replay improvements and low population matching cost.

The population application shows why national agreement is insufficient on its own. RHDM and BSLM differ by only \updated{0.08\%} in national daily trip count, while purpose-specific differences are larger locally and can change direction. Shopping visits increase nationally but decrease among central Geneva residents; differences across residential cells are larger still. The methods should therefore not be regarded as interchangeable simply because their aggregate demand looks similar.

For planning and policy analysis, matching should be evaluated as part of the substantive demand model. Individual diary diagnostics, local sensitivity and downstream transport validation provide complementary evidence. The results show that diary assignment can materially alter local demand inputs and may consequently affect decision making. Establishing the size and direction of actual forecast or policy effects requires the next step: comparing the resulting location choices, simulated travel and policy outcomes against independent evidence.

\section*{Data and code availability}
The data used in this study are subject to the access conditions of the respective data providers in Switzerland and cannot be shared publicly. The code, however, is openly available as part of the eqasim-switzerland pipeline\footnote{\url{https://github.com/eqasim-org/eqasim-switzerland}}. The implementation of the proposed matching algorithm is provided in the file \texttt{synthesis/population/matching/matched\_v3.py}\footnote{\url{https://github.com/eqasim-org/eqasim-switzerland/blob/cmdp_new_version/synthesis/population/matching/matched_v2.py}}.
\FloatBarrier

\FloatBarrier
\bibliographystyle{elsarticle-harv}
{\small\bibliography{references}}

\newpage
\appendix
\section{Technical model details}
\label{sec:technical}
The main text describes the model's purpose and training process. This appendix records the numerical architecture, loss functions and donor-selection formula for readers who wish to reproduce the method.

Categorical variables are organised into residential context, mobility resources, work/education and personal context. Separate fully connected projections reduce these groups to 8, 4, 6 and 8 values. Each projection uses layer normalisation and a sigmoid-weighted linear unit (SiLU), which provide normalisation and a smooth nonlinear transformation \citep{ba2016layernorm,elfwing2018silu}. The projections are joined with 11 scaled numerical values and 11 missingness indicators, giving 48 encoder inputs. A fully connected layer maps these to 64 values, followed by layer normalisation, SiLU and dropout with probability 0.2 \citep{srivastava2014dropout}. A second layer maps the result to the 16-dimensional representation, with layer normalisation.

Gradient norms are clipped at 5 to limit unusually large parameter updates \citep{pascanu2013clipping}. Network training uses PyTorch, while feature preparation and nearest-neighbour searches use scikit-learn \citep{paszke2019pytorch,pedregosa2011sklearn}.

\subsection{Training losses}
Activity purposes and modes use cross-entropy loss, the negative log probability given to the observed category. Numerical outcomes use the Huber loss \citep{huber1964robust},
\begin{equation}
H(e)=\begin{cases}\tfrac12 e^2,& |e|\leq1,\\ |e|-\tfrac12,& |e|>1,\end{cases}
\end{equation}
which is quadratic for small errors and linear for large errors. Each sequence loss is averaged over its observed positions. Missing measurements and padding do not contribute. Departure times are divided by 30 hours. Trip count, distance and duration targets use $\log(1+x)/\log(1+s)$, with reference scales $s$ of five trips, 30\,000\,m and 3\,600\,s, respectively. These scales make outcomes comparable for training; they do not remove larger observations. Up to 12 trips per diary supply sequence-training targets, while the count target and transferred diary retain the full day. In this experiment, 68 diaries exceed that sequence limit.

Let $\ell_{jh}$ denote the loss of person $j$ for behavioural output $h$, and let
$\bar{w}_j = w_j / \left(\frac{1}{|\mathcal{D}|}\sum_{k \in \mathcal{D}} w_k\right)$
be the survey weight $w_j$ of person $j$, normalised by the mean weight of the eligible donors $\mathcal{D}$.
The latent representation has dimension $d = 16$.
For a minibatch $B$, the training objective is
\begin{align}
\mathcal{L}_B ={}& \frac{1}{|B|} \sum_{j \in B} \bar{w}_j \left[ \sum_{h} \lambda_h \, \ell_{jh}
  + 0.2 \, \ell_{j,\mathrm{rec}} + 0.05 \, \frac{\|\tilde{z}_j\|^2}{d} \right] \nonumber \\
  & + 0.2 \, \max\!\left( 0,\; 0.5 - \frac{1}{d} \sum_{r=1}^{d} \operatorname{Var}_B(\tilde{z}_{jr}) \right),
\label{eq:loss}
\end{align}
where $\operatorname{Var}_B(\cdot)$ is the variance of a latent dimension over the minibatch.
The behavioural weights $\lambda_h$ are set to $2$ for activity purposes, $1.5$ for the number of trips,
$0.1$ for modes, $0.1$ for departure times, $0.3$ for trip distances, and $0.5$ each for travel durations and activity durations.

The reconstruction loss $\ell_{j,\mathrm{rec}}$ is the mean squared error with respect to the encoder inputs,
which are treated as constant targets, i.e., no gradient is propagated through them.
This branch follows the autoencoder principle \citep{hinton2006reduction}.
The squared-norm term limits large latent values, while the hinge term on the minibatch variance
discourages a collapse to a constant representation.
The latter is related to the variance regularisation used against representation collapse,
but it is not the full VICReg objective \citep{bardes2022vicreg}.

During training, the latent vector is perturbed as $\tilde{z} = z + \epsilon$,
where $\epsilon \sim \mathcal{N}(0, \sigma^2 I)$ with $\sigma = 0.08$,
which acts as an additional regulariser \citep{bishop1995noise}.
No noise is added when the representation is used for matching, i.e., $z$ is used directly.

\subsection{Donor-selection probabilities}
\label{sec:donor_selection}

Donors (survey respondents) and recipients (synthetic persons) are both encoded with the fitted encoder, and the resulting latent vectors are normalised to unit length, $u_i = z_i / \|z_i\|_2$.
Let $\mathcal{C}_i$ denote the set of donors that are compatible with recipient $i$ (see below).
Among the compatible donors, the $k = 200$ nearest neighbours of $u_i$ in Euclidean distance are retained and form the candidate set $\mathcal{N}_i \subseteq \mathcal{C}_i$; if fewer than $k$ compatible donors exist, all of them are retained, i.e., $|\mathcal{N}_i| = \min(k, |\mathcal{C}_i|)$.
The donor $J_i$ of recipient $i$ is then drawn from $\mathcal{N}_i$ with probability
\begin{equation}
  P(J_i = j \mid x_i) =
  \frac{w_j \exp\!\left[ -\|u_i - u_j\|_2^2 / (2\tau) \right]}
       {\sum_{q \in \mathcal{N}_i} w_q \exp\!\left[ -\|u_i - u_q\|_2^2 / (2\tau) \right]},
  \qquad j \in \mathcal{N}_i,
  \label{eq:donor}
\end{equation}
where $x_i$ denotes the attributes of recipient $i$, $w_j$ is the survey weight of donor $j$, and $\tau = 0.1$ is a temperature parameter.
A smaller value of $\tau$ concentrates the draw on closer neighbours, whereas the survey weights account for the unequal representation of donors among the candidates.
The procedure thus combines nearest-neighbour retrieval \citep{cover1967nearest} with a weighted probabilistic donor draw.
Once a donor has been selected, the complete diary of the donor is transferred to the recipient unchanged at the diary-matching step.

\section{Scoring definitions}
\label{sec:scores}
A single person may plausibly have several different days. We therefore evaluate the empirical distribution of donor draws using proper scoring rules \citep{gneiting2007scores}. All scores below are better when lower. With $M=100$ predicted daily vectors $v_i^{(m)}$ and observed vector $v_i$, the energy score is
\begin{equation}
\operatorname{ES}_i=\frac1M\sum_m\|v_i^{(m)}-v_i\|_2
 -\frac1{2M^2}\sum_{m,q}\|v_i^{(m)}-v_i^{(q)}\|_2.
\label{eq:energy}
\end{equation}
The first term penalises predictions far from the observation; the second accounts for the distribution of predicted days. The vector contains trip count, travel time, daily distance, first departure, last arrival, counts and durations for five non-home purposes, and counts for five travel modes. Components are divided by survey-weighted standard deviations from the available donor groups so that kilometres do not dominate counts. A zero-variance component uses scale one.

For a scalar outcome $a_i$, the continuous ranked probability score (CRPS) replaces Euclidean distance in Equation~\ref{eq:energy} by absolute difference:
\begin{equation}
\operatorname{CRPS}_i=\frac1M\sum_m|a_i^{(m)}-a_i|
 -\frac1{2M^2}\sum_{m,q}|a_i^{(m)}-a_i^{(q)}|.
\end{equation}
It retains the outcome's units. Distance CRPS is expressed in kilometres, travel-time CRPS in minutes, and shopping-count CRPS in visits per day. Both equations use the empirical-distribution convention with all $M^2$ pairs, including identical draws. Ensemble size is the same for every method.

For an event such as shopping, the predicted probability is the fraction of assigned diaries containing that event. Let $p_i$ be this probability for respondent $i$, and let $o_i$ be one if the event occurs in their observed diary and zero otherwise. The event Brier score is
{\color{black}\begin{equation}
\operatorname{BS}_i=(p_i-o_i)^2.
\label{eq:event_brier}
\end{equation}}
This score penalises a confident wrong prediction more strongly. For example, predicting a shopping probability of 0.3 gives a score of 0.49 if shopping is observed and 0.09 if it is not. The score lies between zero and one; zero means perfect agreement \citep{brier1950verification}.

For a complete ordered activity-and-mode chain, there are several possible outcomes rather than a single yes/no event. Let $p_{is}$ be the fraction of assigned diaries with chain $s$, and $c_i$ the observed chain. The chain Brier score adds the squared probability errors over the distinct observed or assigned chains $\mathcal S$:
{\color{black}\begin{equation}
\operatorname{CBS}_i=\sum_{s\in\mathcal S}\left[p_{is}-\mathbf{1}\{c_i=s\}\right]^2.
\label{eq:chain_brier}
\end{equation}}
The indicator is one for the observed chain and zero for the others. A low score therefore requires both a high probability for the observed chain and little probability assigned to alternatives. This is the summed multi-category Brier score, without dividing by the number of chains; it ranges from zero to two.

The variogram score checks whether relationships between daily quantities are also reproduced \citep{scheuerer2015variogram}. Using daily quantities $v_i$ scaled by their donor-group standard deviations, without subtracting their means, and assigned vectors $v_i^{(m)}$, it compares their pairwise differences:
{\color{black}\begin{equation}
\operatorname{VS}_i=\frac{1}{|\mathcal P|}\sum_{(a,b)\in\mathcal P}\left[|v_{ia}-v_{ib}|^{1/2}-\frac1M\sum_{m=1}^M|v_{ia}^{(m)}-v_{ib}^{(m)}|^{1/2}\right]^2.
\label{eq:variogram}
\end{equation}}
Here $M=100$, and $\mathcal P$ contains five pairs: trip count with travel time, distance and first departure; travel time with distance; and first departure with last arrival. This adds a check of the joint daily pattern, beyond scoring each quantity separately. Standardisation prevents different units from dominating the comparison.

Mean replay scores use survey weights. Comparisons use exactly the same recipients. We obtain 95\% intervals by resampling households 500 times and calculate method differences within each resample \citep{davison1997bootstrap}. These intervals describe respondent-sampling uncertainty conditional on the fitted encoder, donor partition and empirical draws. They do not include uncertainty from retraining, survey selection or alternative hyperparameters.
\section{A visual view of the learned representation}
\label{sec:latent}
\begin{figure}[htbp]
\centering\includegraphics[width=.90\textwidth]{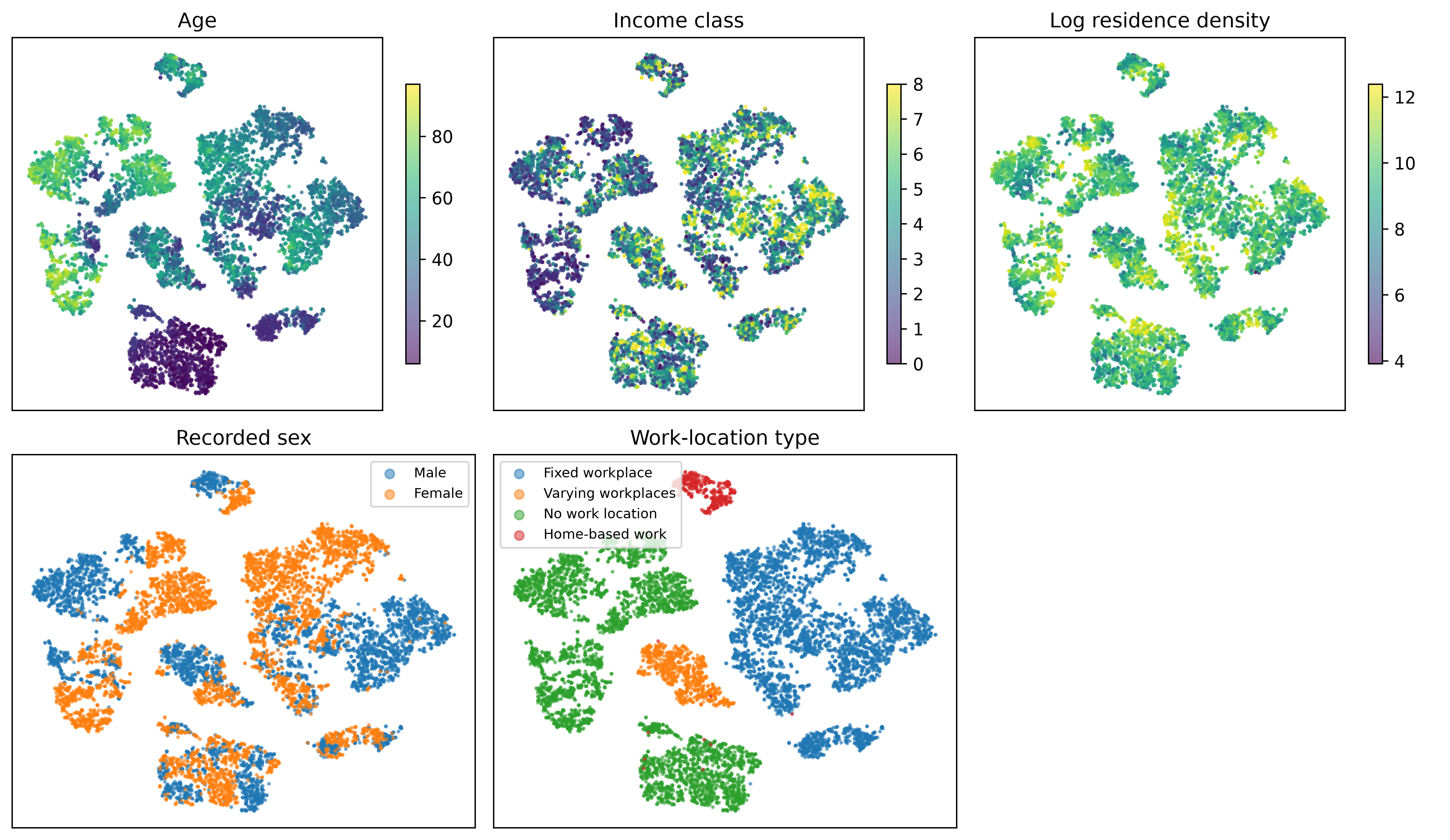}
\caption{Nonlinear display of the learned representation, coloured by exact age, original income class, log residence density, recorded sex and work-location type. All panels use identical t-SNE coordinates. Axes have no physical units.}
\label{fig:latent}
\end{figure}

\autoref{fig:latent} provides a qualitative view of the encoder used in the reported experiments. The same 10\,000 randomly selected eligible respondents appear in every panel. A t-distributed stochastic neighbour embedding (t-SNE) projection reduces their 16-dimensional normalised vectors to two display coordinates \citep{vandermaaten2008tsne}. Perplexity is 30 and the random seed is fixed. Colours describe inputs rather than predicted behavioural classes. The display can help inspect structure, but apparent clusters, their sizes and separation are not direct measurements of behavioural similarity or evidence of causation \citep{wattenberg2016tsne}. 

\end{document}